\documentclass{article}
\pdfoutput=1
\usepackage{jheppub}
\usepackage{tensor}
\usepackage{physics}

\title{Point-Particle Catalysis}

\author[a,b]{P.~Hayman}
\emailAdd{haymanpf@mcmaster.ca}
\author[a,b]{and C.P.~Burgess}
\emailAdd{cburgess@perimeterinstitute.ca} 
\affiliation[a]{Physics \& Astronomy, McMaster University, Hamilton, ON, Canada, L8S 4M1}
\affiliation[b]{Perimeter Institute for Theoretical Physics, Waterloo, Ontario N2L 2Y5, Canada }

\date{\today}

\abstract
{
	We use the point-particle effective field theory (PPEFT) framework to describe particle-conversion mediated by a flavour-changing coupling to a point-particle. We do this for a toy model of two non-relativistic scalars coupled to the same point-particle, on which there is a flavour-violating coupling. It is found that the point-particle couplings all must be renormalized with respect to a radial cut-off near the origin, and it is an invariant of the flow of the flavour-changing coupling that is directly related to particle-changing cross-sections. At the same time, we find an interesting dependence of those cross-sections on the ratio $k_{\text{out}}/k_{\text{in}}$ of the outgoing and incoming momenta, which can lead to a $1/k_{\text{in}}$ enhancement in certain regimes. We further connect this model to the case of a single-particle non-self-adjoint (absorptive) PPEFT, as well as to a PPEFT of a single particle coupled to a two-state nucleus. These results could be relevant for future calculations of any more complicated reactions, such as nucleus-induced electron-muon conversions, monopole catalysis of baryon number violation, as well as nuclear transfer reactions.
}

\begin{document}
\maketitle

\section{Introduction}
\label{section:Intro}

It is often the case that physically interesting situations involve a hierarchy of characteristic scales. For instance, solar system dynamics involve a variety of length scales, such as the sizes of the stars and planets involved, as well as the sizes of the orbits. Exploiting such a hierarchy by means of judicious Taylor expansions can greatly simplify otherwise very difficult problems, frequently even providing a handle on seemingly intractable problems. In the realm of quantum field theory, this insight has led to the development of the highly successful effective field theories, which can reduce the complexity of quantum field theories by restricting to parameter subspaces in which an appropriate Taylor expansion can be used to put the theory into a simpler form.

Usually, effective field theories exploit the hierarchy between interaction energies and the masses of some heavy particles to remove those heavy particles from the theory altogether (the quintessential example being Fermi's theory of the Weak interaction, which removes the heavy $W$ and $Z$ bosons) \cite{burgess_introduction_2007,burgess_standard_2006,weinberg_quantum_1995,weinberg_quantum_1996}. However, it is often the case that one's interest lies in a sector of the theory that still contains one or two of the heavy particles. For instance, in an atom, a heavy nucleus is present, but for most purposes there's no need to go about computing loops of nucleus-anti-nucleus pairs. Instead, higher energy nuclear dynamics are seen as finite nuclear-size effects. For this reason, an EFT has recently been explored that describes the remnant heavy particles in position-space to exploit the hierarchy of energy scales in a more intuitive expansion in $kR$, where $k$ is the (small) momentum of the light particle and $R$ is the length-scale of the nuclear structure \cite{burgess_point-particle_2017,burgess_point-particle_2017-1,burgess_point-particle_2017-2}. This is accomplished in a simple way; the usual effective action is supplemented by a ``point-particle'' action that involves all possible couplings of the light particle to the worldline of the remnant heavy (point-like) particle (consistent with the symmetries of the low-energy theory). This type of ``point-particle'' EFT (PPEFT) is conceptually the next best thing to a Fermi type of EFT. While the nuclear dynamics cannot be removed altogether, they are significantly simplified. 

The practicality of a PPEFT is twofold: first it easily permits parameterizing physical quantities in terms of small nuclear properties, since the PPEFT expansion is directly in powers of $kR$. Second, that parameterization is completely general, and inherently includes all possible interactions, including any potential new physics. Some obvious examples that have been explored are cross-sections and bound-state energies of electrons in terms of nuclear charge radii \cite{burgess_point-particle_2017-2,burgess_reduced_2018}. In this work, we ask the question: how do the small nuclear properties enter into physical quantities when there are multiple channels of interaction with the point-particle? (We ask this with mind towards eventually describing nuclear transfer reactions, and possibly baryon-monopole reactions). 

To answer this question, we consider a simple toy model of two bulk Schr\"odinger fields coupled to the same point particle. The most general couplings to the point-particle's worldline $y^\mu(\tau)$ are easily generalized from the single particle species (SP) examples explored in \cite{burgess_point-particle_2017} to the multi-particle (MP) case. 
\begin{equation}
	\label{eq:intro:brane}
	S_b^{(\text{SP})} = -\int \dd \tau \sqrt{-\dot y^2} \, [M + h\, \Psi^* \Psi + \dots] \quad \longrightarrow \quad S_{b}^{(\text{MP})} = -\int \dd \tau \sqrt{-\dot y^2} \, [M + \Psi^*_a h_{ab} \Psi_b + \dots],
\end{equation}
where $\Psi, \,\Psi_a$ are bulk scalars, and the flavour index runs over 1 and 2. $h_{ab}$ is a matrix of coupling constants that generalizes the single-particle coupling $h$, and the integral is over the proper time $\tau$ of the point-particle ($\dot y^{\mu} := \dv[]{y^\mu}{\tau}$ is the point-particle's 4-velocity). Away from the point-particle, the action is just the usual Schr\"odinger action for (now) two scalars, 
\begin{equation}
	\label{eq:intro:bulk}
	S_{B}^{(\text{MP})} = \sum_{a=1}^2 \int \dd^{4}x \, \left\{\frac{i}{2}\left( \Psi^*_a \partial_t \Psi_a - \Psi_a \partial_t \Psi_a^* \right) - \frac{1}{2m_a}\abs{\nabla \Psi_a}^2 - V(r)\abs{\Psi_a}^2\right\}.
\end{equation}
($V(r)$ is some bulk potential that may be sourced by the point-particle. In the main text we take it to be an inverse-square potential, since such a potential is highly singular and known to drive interesting behaviour in a PPEFT \cite{burgess_point-particle_2017}. For the moment it suffices to drop the potential). If the bulk action \eqref{eq:intro:bulk} diagonalizes the momentum it need not diagonalize the brane action \eqref{eq:intro:brane}, and it is possible the off-diagonal elements of $\bf h$ (the matrix of $h_{ij}$) can source flavour-violating interactions. 

In the centre-of-mass reference frame (in the limit of infinite point-particle mass\footnote{We neglect here recoil effects, though those can be included by tracking the dynamics of $y^\mu(\tau)$.}), the action \eqref{eq:intro:brane} acts as a boundary condition at the origin for the modes of the $\Psi_a$ fields. However, in general those diverge, and so the action has to be regulated at some finite radius $\epsilon$. The couplings $\bf h$ must then be renormalized to keep physical quantities independent of the regulator, and it turns out that the (low-energy $s$-wave) cross-section for flavour violation is directly related to an invariant ($\epsilon_3$) of the RG-flow of the off-diagonal $h_{12}$:
\begin{equation}
	\label{eq:intro:catXs}
	\sigma^{(1\to 2)}_s = 4\pi\frac{k_2}{k_1}\epsilon_3^2,
\end{equation}
where $k_1$ and $k_2$ are the incoming and outgoing momenta, respectively. For Schr\"odinger particles, the factor $k_2/k_1 = \sqrt{m_2/m_1} $ is a constant. However the same formula holds for spinless relativistic particles, and the ratio $k_2/k_1 = \sqrt{(k_1^2 + m_1^2 - m_2^2)/k_1^2} $ leads to different qualitative behaviours of the low-energy cross-section depending on how $k_1$ relates to the mass gap $m_1^2 - m_2^2$. If the mass gap is positive, and $k_1^2 \ll m_1^2 - m_2^2$, then the cross-section exhibits a $1/k_1$ enhancement. Both the dependence on $\epsilon_3$ and on $k_1$ may prove to be useful in a more complicated calculation, such as in mesic transfer reactions $\pi^0 + p \to \pi^+ + n$ (where the neutron and proton are in a nucleus), or possible flavour changing reactions involving new physics, such as $\mu^- + N \to e^- + N$ \cite{pezzullo_mu2e_2017,bonventre_searching_2019}.

The channels of interaction with the point-particle do not have to be different bulk species, however. If, for example, the nucleus carried two accessible energy states, say $E_\uparrow = M + \Delta/2$ and $E_\downarrow = M - \Delta/2$ (where $\Delta \ll M$ is some small excitation energy), then two channels of interaction could be a single bulk particle interacting with each of the nuclear energy eigenstates. In this case the ``flavour-violating'' cross section is again \eqref{eq:intro:catXs}, where now $k_1$ and $k_2$ are the incoming and outgoing single-particle momenta, and $k_2/k_1 = \sqrt{(k_1^2 \pm 2m\Delta)/k_1^2}$ with the $\pm$ corresponding to the bulk particle impinging on a nucleus in the ground state ($-$) or the excited state ($+$). On its own, this description is enough for any simple reaction $\psi + N \to \psi + N^*$, where an incident non-relativistic particle just knocks a nucleus into a long-lived slightly energized state. Together, the two-species and two-nuclear state models form the building blocks for exploring more complicated processes, such as nuclear transfer reaction, where an incident particle exchanges some constituent particles with a nucleus, and the final state \emph{both} violates flavour of the bulk species \emph{and} changes the state of the source nucleus.

Finally, we can also look towards simpler models instead. One may imagine for instance only being interested in tracking one of the bulk-species, say particle 1 (perhaps an apparatus can only detect particles of flavour 1). In that case, the flavour-violating cross-section appears as an absorptive interaction when restricting to the particle 1 subspace of the theory. In this way, our toy model can be seen as a particular unitary completion of a model with a single particle subject to a non-self-adjoint Hamiltonian, as studied in \cite{plestid_fall_2018} and frequently used in the form of nuclear optical models \cite{feshbach_nuclear_2003,dickhoff_recent_2019}.  

The rest of this paper is organized as follows. First, we briefly recall the salient details of a simple PPEFT for a single bulk species. Then in \S \ref{section:action}, we establish the action and classical solutions to a point-particle EFT involving two bulk species, followed by \S \ref{section:BCs} in which we solve the boundary conditions of the system, and determine how all of the point-particle couplings run. All of this comes together in \S \ref{section:scattering} where we compute how the point-particle properties relate to physical cross-sections, including the cross-section for flavour-violation. In \S \ref{section:rel:multiNuc} we connect the multi-bulk species story to a single particle coupled to a two-state nucleus. Finally, we wrap up in \S \ref{section:rel:abs} by restricting to a single-particle subsector of the multi-species model, and realizing the equivalence to the absorptive model of \cite{plestid_fall_2018}.

\section{Point-Particle EFT for a Single Bulk Species}
\label{section:review}

We review the point-particle effective field theory for a single Schr\"odinger particle in an inverse-square potential, first described in \cite{burgess_point-particle_2017}. 

In the point-particle effective field theory approach, we exploit the hierarchy of length-scales between the characteristic wavelength of some low-energy particle of mass $m$ (for concreteness, call this some scalar electron) and the scale of some small, almost point-like particle of mass $M \gg m$ it interacts with (similarly, we'll call this a nucleus). For example, in atomic systems, this would be the ratio $R/a_0$ between the size $R$ of a nucleus and the Bohr radius $a_0$ of the atom. For scattering, the small parameter is more directly $kR$, with $k$ the wavenumber of the incident particle. The way we exploit this hierarchy is to recall that the low energy dynamics of the heavy particle are well approximated by ordinary quantum mechanics, so we imagine only first-quantizing the nucleus. In that case the fully second-quantized electron only couples to the 1-dimensional world-line of the heavy particle. This amounts to writing the action for the electron $S = S_B + S_b$ in terms of the usual bulk dynamics\footnote{We use a mostly plus metric, and work in units such that $\hbar = c = 1$.}
\begin{equation}
	\label{eq:review:bulkAction}
	S_{B} = \int \dd^{4}x \, \left\{\frac{i}{2}\left( \Psi^* \partial_t \Psi - \Psi \partial_t \Psi^* \right) - \frac{1}{2m}\abs{\nabla \Psi}^2 - V(r)\abs{\Psi}^2\right\}
\end{equation}
as well as a boundary term consisting of interactions between the electron and the nuclear worldline,
\begin{equation}
	\label{eq:review:PPaction}
	S_{b} = -\int \dd \tau \sqrt{-\dot y^2}\left(M + h\, \abs{\Psi(y)}^2 + \dots \right).
\end{equation}
In \eqref{eq:review:bulkAction}, $V(r)$ may be some potential sourced by the point-particle, and the dots represent terms of higher powers in $kR$. In what follows, we choose $V(r) = \frac{g}{r^2}$, since an inverse-square potential is known to be responsible for interesting non-trivial behaviour in a PPEFT \cite{burgess_point-particle_2017,burgess_point-particle_2017-1}. Now through \eqref{eq:review:PPaction}, there are couplings on the world-line $y^{\mu}(\tau)$ of the nucleus (parameterized by its proper time $\tau$, so that $\dot y^\mu := \dv{y^\mu}{\tau}$ is the 4-velocity of the nucleus). The first term $\sqrt{-\dot y^2}  M$ can be recognized as the usual action for a point-particle \cite{lambert_string_2012}, while the second term is the lowest-order (in powers of length) coupling between the electron and the nucleus, with the dots representing interactions of higher order in $kR$. For a spherically symmetric nucleus, the coupling $h$ is a constant.

For simplicity, and to emphasize the value of the point-particle interactions, we work in the limit of infinite nuclear mass, where $y^\mu = (t, 0, 0, 0)$ is the centre-of-mass frame, and $\tau = t$. This amounts to neglecting nuclear recoil, but that can be included by explicitly tracking the dynamics of $y^\mu(\tau)$. Variation of the total action $S$ with respect to $\Psi^{*}$ yields the usual Schr\"odinger equation in the bulk 
\begin{equation}
	\label{eq:review:bulkEOM}
	\left(i\partial_t - \frac{\nabla^2}{2m} - \frac{g}{r^2}\right)\Psi = 0,
\end{equation}
as well as the boundary condition 
\begin{equation}
	\label{eq:review:BC}
	\lim_{\epsilon \to 0}\, 4\pi\epsilon^2 \partial_\epsilon \Psi_\ell = \lim_{\epsilon \to 0}\, 2mh_\ell \Psi_\ell(\epsilon),
\end{equation}
which defines $\partial_\epsilon := \partial_r\vert_\epsilon$. In \eqref{eq:review:BC}, $\Psi_\ell$ is the $\ell^\text{th}$ eigenfunction of angular momentum, and the boundary condition is evaluated on a sphere of radius $\epsilon$ (in the limit $\epsilon \to 0$) in case the wavefunction or its derivative is not finite in that limit. If $\epsilon^2\partial_\epsilon \Psi$ is finite, then the limit $\epsilon \to 0$ is consistent with $h_\ell = 0$ (The fact that $h$ couples differently to each $\ell$ mode can be thought of as resolving the coordinate ambiguity arising from extrapolating $r = 0$ to a non-zero value\footnote{More correctly, imagine a UV completion with an interaction $\int \dd^4 x \, h_{UV} \abs{\Psi}^2\abs{\chi}^2$, where $\chi$ is the nuclear field. The PPEFT limit is obtained by perturbing around energies just larger than $M$, so that the dominant behaviour of $\chi$ is captured by a wavepacket solution centred at $y^\mu(\tau)$. In that regime, the point-particle action arises from performing the spatial integral over a ball $\mathcal{B}_\epsilon$ centred at $y^i(\tau)$, and using that the electron wavefunction in this regime is roughly constant. The matching to the effective coupling $h$ is essentially $h = h_{UV}\int_{\mathcal{B}_\epsilon}\dd^3 x \,\abs{\chi}^2$. The range of validity of the approximation that $\abs{\Psi}^2$ is constant is an $\ell$-dependent condition, and so the different $h_\ell$ arise as different limits of the integral of the nuclear field.}). If either $\Psi_\ell$ or its derivative is \emph{not} finite in the limit $\epsilon \to 0$, then the boundary condition \eqref{eq:review:BC} establishes first \emph{that} $h_\ell$ must be renormalized and secondly \emph{how} it must run with $\epsilon$ in order to keep physical quantities independent of the regulator.

The bulk equations \eqref{eq:review:bulkEOM} are solved by 
\begin{equation}
	\label{eq:review:bulkSol}
	\Psi = e^{-iEt}\left( C_+ \psi_+ + C_- \psi_- \right), 
\end{equation}
where $E$ is the electron energy, and the mode functions are
\begin{equation}
	\label{eq:review:modes}
	\psi_{\pm}(\rho) := \rho^{\frac{1}{2}(-1\pm\zeta)}e^{-\rho/2}\mathcal{M} \left[ \frac{1}{2}\left( 1\pm\zeta \right),1\pm\zeta;\rho \right]
\end{equation}
which defines $k^2 := 2mE$, $\rho := 2ikr$, and $\zeta := \sqrt{(2\ell + 1)^2 - 8mg}$. For simplicity, in this paper we will restrict to the case $mg \leq 1/8$ so that $\zeta$ is always real. Taking the small $k\epsilon$ limit of \eqref{eq:review:modes}, the boundary condition determines the renormalization-group flow of the coupling $h_\ell$ through
\begin{equation}
	\label{eq:review:run}
	\hat \lambda = \frac{1 - \frac{C_-}{C_+}(2ik\epsilon)^{-\zeta}}{1 + \frac{C_-}{C_+}(2ik\epsilon)^{-\zeta}} = \frac{1 + (\epsilon/\epsilon_\star)^{-\zeta}}{1 - (\epsilon/\epsilon_\star)^{-\zeta}},
\end{equation}
where $\hat \lambda := \frac{1}{\zeta}(mh/\pi\epsilon + 1)$ (we drop the subscript $\ell$ for convenience), $y := \text{sgn}(\abs*{\hat\lambda} - 1)$ defines a renormalization-group trajectory, and $\epsilon_\star$ is an RG-invariant length scale, both determined by the physical quantity
\begin{equation}
	\label{eq:review:defEps}
	\frac{C_-}{C_+} = -y(2ik\epsilon_\star)^{\zeta}.
\end{equation}

Physical quantities like scattering cross-sections and bound-state energies are directly related to the ratio $C_-/C_+$, and so through \eqref{eq:review:defEps} directly to the quantity $\epsilon_\star$, which is fundamentally a property of the source. The usefulness of the inverse-square potential lies in how it can force non-trivial RG behaviour upon the point-particle coupling. For example, the running \eqref{eq:review:run} has an ``infrared'' fixed point of $+1$ when $\epsilon/\epsilon_\star \to \infty$, which corresponds to $\epsilon_\star \to 0$. For the $s$-wave in the absence of an inverse-square potential, $\zeta(\ell = 0) = 1$ and this would be equivalent to vanishing point-particle coupling, but if the strength of the inverse-square potential $g \neq 0$, then the fixed point is driven away from a vanishing point-particle coupling.

In the next section, we generalize all of this to a bulk system composed of multiple species of particles (though for concreteness, we specialize to two species). We see how most of the above follows through identically, but the presence of boundary terms that mix flavours adds a new degree of complexity to the problem, introducing a new point-particle coupling which runs differently from \eqref{eq:review:run} and opening the door to flavour-changing reactions.

\section{Multi-Species Action and Bulk Field Equations}
\label{section:action}

The simplest extension of the basic point-particle action \eqref{eq:review:PPaction} to multiple particles is a non-diagonal quadratic one:
\begin{equation}
	\label{eq:action:PPaction}
	S_{b}^{(\text{MP})} = -\int \dd \tau \sqrt{-\dot y^2} \, [M + \Psi^*_a h_{ab} \Psi_b + \dots]
\end{equation}
where now there are $N$ complex scalar Schro\"odinger fields $\Psi_i$, and summation over the species index is implied. The bulk action is taken to diagonalize the momentum operator, so flavour mixing only happens at the point particle, and the bulk action is simply $N$ copies of \eqref{eq:review:bulkAction}:
\begin{equation}
	\label{eq:action:bulkAction}
	S_{B}^{(\text{MP})} = \sum_{a=1}^N \int \dd^{4}x \, \left\{\frac{i}{2}\left( \Psi^*_a \partial_t \Psi_a - \Psi_a \partial_t \Psi_a^* \right) - \frac{1}{2m_a}\abs{\nabla \Psi_a}^2 - V(r)\abs{\Psi_a}^2\right\}.
\end{equation}

For concreteness, we will work with only $N = 2$ species of particles. Our interest is in single particle states, so we restrict to the single-particle sector, for which the Hilbert space is $\mathcal{H} = \mathbb{C}\oplus\mathcal{H}_1\oplus\mathcal{H}_2$ (where $\mathcal{H}_i$ is the Hilbert space for particle $i$). On this space, the Schr\"odinger operator $i\partial_t - \frac{\nabla^2}{2m_i} - \frac{g}{r^2}$ is diagonal. Writing the total wavefunction $\Psi = \mqty(\Psi_1 \\ \Psi_2)$, the equations of motion read
\begin{equation}
	\label{eq:action:TDEOM}
	\mqty[i\partial_t - \frac{\nabla^2}{2m_1} - \frac{g}{r^2} & 0 \\
	0 & i\partial_t - \frac{\nabla^2}{2m_2} - \frac{g}{r^2} ]\mqty[\Psi_1 \\ \Psi_2] = 0.
\end{equation}
The time-dependence is easily solved using separation-of-variables: $\Psi = e^{-iE t}\mqty(\psi_1 \\ \psi_2)$. Then we find
\begin{equation}
	\label{eq:action:EOM}
	\mqty[k_1^2 + \nabla^2 - \frac{g}{r^2} & 0 \\
	0 & k_2^2 + \nabla^2 - \frac{g}{r^2}]\mqty[\psi_1 \\ \psi_2] = 0,
\end{equation}
which defines the wavenumbers $k_i := \sqrt{2m_iE} $ (real for continuum states and imaginary for bound states).

Away from the origin, \eqref{eq:action:EOM} is solved exactly as in the single-species problem, and a natural choice of basis for the solutions is
\begin{equation}
	\label{eq:action:bulkBasis}
	\mathcal{B} = \qty{\mqty(\psi_1 \\ 0), \, \mqty(0 \\ \psi_2)},
\end{equation}
where
\begin{equation}
	\label{eq:action:bulkFuncs}
	\psi_i = \left(C_{i+}\psi_{+}(\ell_i; 2ik_ir) + C_{i-}\psi_{-}(\ell_i; 2k_{i}r)\right)Y_{\ell_i m_i}.
\end{equation}
and
\begin{equation}
	\label{eq:action:bulkModes}
	\psi_{\pm}(\ell;\rho) := \rho^{\frac{1}{2}(-1\pm\zeta_i)}e^{-\rho/2}\mathcal{M} \left[ \frac{1}{2}\left( 1\pm\zeta_i \right),1\pm\zeta_i;\rho \right]
\end{equation}
with $\rho := 2ik r$, and $\zeta_i := \sqrt{(2\ell+1)^2 - 8m_ig}$, as in the single-particle example.

The constants $C_{i\pm}$ are solved by considering the boundary conditions in the problem, typically finiteness at large- and small-$r$, but for scattering problems the large-$r$ BC is specific to the setup (since it depends on the presence or otherwise of incident particles). In a PPEFT, the small-$r$ boundary condition is derived from the boundary action, which we describe next.  

\section{Boundary Conditions}
\label{section:BCs}

In analogy with the single-particle system, we determine the small-$r$ boundary condition by varying the point-particle action \eqref{eq:action:PPaction} directly (including the boundary terms in the variation of the bulk action \eqref{eq:action:bulkAction}). The resulting small-$r$ boundary conditions are a simple generalization of \eqref{eq:review:BC}: 
\begin{equation}
	\label{eq:BC:genBC}
	\lim_{\epsilon\to 0}\, 4\pi\epsilon^2\partial_\epsilon \Psi = \lim_{\epsilon\to 0}\, 2\mathbf{mh}\Psi(\epsilon),
\end{equation}
again using $\Psi = e^{-iE t}\mqty(\psi_1 \\ \psi_2)$ and as in section \ref{section:review}, we define $\partial_\epsilon := \partial_r\vert_\epsilon$. Here $\bf{m}$ and $\bf{h}$ are the mass and point-particle coupling matrices (respectively), so that in components (the limit as $\epsilon \to 0$ implied),
\begin{subequations}
	\label{eq:BC:SmallRBC}
\begin{align}
	4\pi\epsilon^2\partial_\epsilon\psi_1 - 2m_1h_{11}\psi_1(\epsilon) - 2m_1h_{12}\psi_{2}(\epsilon)
 &= 0 \qquad \text{and} \label{eq:SmallRBC:1}  \\
	4\pi\epsilon^2\partial_\epsilon\psi_2 - 2m_2h_{22}\psi_2(\epsilon) - 2m_2h_{21}\psi_{1}(\epsilon) &= 0. \label{eq:SmallRBC:2}
\end{align}
\end{subequations}
Notice that the explicit $\epsilon$-dependence of \eqref{eq:BC:SmallRBC} again indicates that the point-particle couplings $h_{ij}$ \emph{must} be renormalized with $\epsilon$ whenever $\Psi$ or $\partial_\epsilon \Psi$ diverge for small argument, in order for the boundary condition to be compatible with the bulk equations of motion. 

The boundary condition then serves two distinct purposes: \textbf{i)} solving for the integration constants in $\Psi$ tells us how they (and correspondingly physical quantities) depend on the point-source physics, and \textbf{ii)} isolating for the couplings $h_{ij}$ then tells us how exactly each coupling flows with $\epsilon$ to ensure the physical integration constants do not. Clearly though, with 4 possible degrees of freedom in $\bf h$ and 4 integration constants in $\Psi$, the 2 equations in \eqref{eq:BC:SmallRBC} are insufficient by themselves. In the next sections we invoke physical arguments to resolve this predicament, and separately tackle both problems \textbf{i)} and \textbf{ii)}.

\subsection{Solving for Integration Constants}
\label{section:BCs:IC}

The most obvious place to look for additional constraints is at another boundary. In spherical coordinates, this amounts to looking at the asymptotic behaviour of $\Psi$ as $r \to \infty$. However the asymptotic behaviour of the system is not unique, and is a very situation-dependent property. Since our interest in this paper is in catalysis of flavour violation, it is most pertinent to study scattering states.  

First, focus on scattering $\Psi_1 \to \Psi_j$. In this case, asymptotically we need (see appendix \ref{appendix:scattering:Multi} for a review of multi-species scattering) $\psi_1(r \to \infty) \to  e^{ik_1z} + f_{11}(\theta,\psi)\frac{e^{ik_1r}}{k_1 r}$ as usual, and now also $\psi_2(r \to \infty) \to f_{12}(\theta,\psi)\frac{e^{ik_2 r}}{k_2 r}$. Notice that both boundary conditions and the equations of motion are linear in $\Psi$, so we may divide through by one integration constant. For incident particle 1, we'll choose to divide through by $C_{1+}$, and we'll define
\begin{equation}
	\label{eq:BC:IC:defICA}
	C_{11} := \frac{C_{1-}}{C_{1+}}, \quad \text{and} \quad C_{12} := \frac{C_{2-}}{C_{1+}} \qquad (1\to X \text{ scattering}),
\end{equation}
and eliminating the infalling wave in $\psi_2$ fixes $C_{2+} = R\,C_{2-}$ with
\begin{equation}
	\label{eq:BC:IC:defR}
	R := - \frac{\Gamma\left( 1 - \zeta/2 \right)}{\Gamma\left( 1 + \zeta/2 \right) }2^{-2\zeta}e^{-i\pi\zeta}
\end{equation}
As we will see in the section \ref{section:scattering}, $C_{11}$ and $C_{12}$ are directly related to the physical cross-sections for $\Psi_1 \to \Psi_1$ and $\Psi_1 \to \Psi_2$ scattering.

Using the definitions \eqref{eq:BC:IC:defICA} and \eqref{eq:BC:IC:defR} in the small-$r$ boundary condition \eqref{eq:BC:SmallRBC} yields
\begin{subequations}
	\label{eq:BC:IC:rbcA}
\begin{align}
	4\pi\epsilon^2\partial_\epsilon\left( \psi_{1+} + C_{11}\psi_{1-} \right) - h_{11}\left( \psi_{1+} + C_{11}\psi_{1-} \right) - h_{12} C_{12}\left( R\,\psi_{2+} + \psi_{2-} \right) 
 &= 0 \qquad \text{and} \label{eq:BC:IC:rbcA:1}  \\
	4\pi\epsilon^2 C_{12}\partial_\epsilon\left( R\,\psi_{2+} + \psi_{2-} \right) - h_{22}C_{12}\left( R\,\psi_{2+} + \psi_{2-} \right) - h_{21}\left( \psi_{1+} + C_{11}\psi_{1-} \right)  &= 0, \label{eq:BC:IC:rbcA:2}
\end{align}
\end{subequations}
Finally, in terms of integration constants the boundary condition \eqref{eq:BC:IC:rbcA} is now a system of two equations for two unknowns, so using the small-$r$ forms of the bulk modes \eqref{eq:action:bulkModes}, 
\begin{equation}
	\label{eq:BC:IC:bulkModesLO}
	\psi_{\pm}(\ell;\epsilon) \approx (2ik\epsilon)^{\frac{1}{2}(-1\pm\zeta)} \qquad \implies \qquad \eval{\pdv{}{r}}_\epsilon\psi_{\pm}(\ell;\rho) \approx ik(-1\pm\zeta)(2ik\epsilon)^{\frac{1}{2}(-3\pm\zeta)},
\end{equation}
it is found (see appendix \ref{appendix:2PICs} for details of the calculation) that the integration constants for this system are related to the source physics through
\begin{equation}
	\label{eq:BC:IC:solveCsLOA}
	\begin{aligned}
		C_{11} = -(2ik_1\epsilon)^{\zeta_1} \,\frac{\widehat{\mathcal{N}}_1}{\widehat{\mathcal{D}}} \qquad &\text{and} \qquad C_{12} = (2ik_1\epsilon)^{\zeta_1/2}(2ik_2\epsilon)^{\zeta_2/2}\sqrt{\frac{m_2k_2\zeta_1}{m_1k_1\zeta_2}}\frac{\hat{\lambda}_{21}}{\widehat{\mathcal{D}}}, 
	\end{aligned}
\end{equation}
where
\begin{equation}
	\label{eq:BC:defNDA}
	\begin{aligned}
	\widehat{\mathcal{N}}_1 &:= 4\widehat{\lambda}_{12}\widehat{\lambda}_{21} - \left[\widehat{\lambda}_{11} - 1 \right]\left[\widehat{\lambda}_{22} + 1\right] 
	\end{aligned}, \qquad \text{and} \qquad 
	\widehat{\mathcal{D}} := 4\widehat{\lambda}_{12}\widehat{\lambda}_{21} - \left[\widehat{\lambda}_{11} + 1 \right]\left[\widehat{\lambda}_{22} + 1\right], 
\end{equation}
and the following convenient re-definitions have been made:
\begin{equation}
	\label{eq:BC:defHats}
	\begin{split}
	\widehat{\lambda}_{11} := \frac{1}{\zeta_1}\left( \frac{m_1h_{11}}{\pi\epsilon} + 1 \right), &\qquad \widehat{\lambda}_{22} := \frac{1}{\zeta_2}\left( \frac{m_2h_{22}}{\pi\epsilon} + 1 \right), \qquad \text{and}, \\ 
	\widehat{\lambda}_{12} := \frac{h_{12}\sqrt{m_1m_2} }{2\pi\epsilon\sqrt{\zeta_1\zeta_2} }, &\qquad \widehat{\lambda}_{21} := \frac{h_{21}\sqrt{m_1m_2} }{2\pi\epsilon\sqrt{\zeta_1 \zeta_2}}.
	\end{split}
\end{equation}

Of course, the choice to make particle 1 the incident particle was not forced upon us, and with foresight to the next sections, we also compute the quantities involved in scattering $\Psi_2 \to \Psi_i$. Fortunately, this is exactly the $1 \leftrightarrow 2$ inversion of the $1\to X$ scattering above, so we can immediately write:
\begin{subequations}
	\label{eq:BC:IC:rbcB}
\begin{align}
	4\pi\epsilon^2 C_{12}\partial_\epsilon\left( R\phi_{2+} + \phi_{2-} \right) - h_{22}C_{12}\left( R\phi_{2+} + \phi_{2-} \right) - h_{21}\left( \phi_{1+} + C_{11}\phi_{1-} \right)  &= 0 \qquad \text{and} \label{eq:BC:IC:rbcB:1}  \\
 \label{eq:BC:IC:rbcB:2}
	4\pi\epsilon^2\partial_\epsilon\left( \phi_{1+} + C_{11}\phi_{1-} \right) - h_{11}\left( \phi_{1+} + C_{11}\phi_{1-} \right) - h_{12} C_{12}\left( R\phi_{2+} + \phi_{2-} \right) 
 &= 0,
\end{align}
\end{subequations}
where as above we've defined
\begin{equation}
	\label{eq:BC:IC:defICB}
	C_{22} := \frac{C_{2-}}{C_{2+}}, \quad \text{and} \quad C_{21} := \frac{C_{1-}}{C_{2+}} \qquad (2\to X \text{ scattering}),
\end{equation}
now with $C_{1+} = R\,C_{1-}$. Solving for the integration constants similarly yields
\begin{equation}
	\label{eq:BC:IC:solveCsLOB}
	\begin{aligned}
		C_{22} = -(2ik_2\epsilon)^{\zeta_2} \frac{\widehat{\mathcal{N}}_2}{\widehat{\mathcal{D}}} \qquad &\text{and} \qquad C_{21} = (2ik_1\epsilon)^{\zeta_1/2}(2ik_2\epsilon)^{\zeta_2/2}\sqrt{\frac{m_1k_1\zeta_2}{m_2k_2\zeta_1}}\frac{\hat{\lambda}_{12}}{\widehat{\mathcal{D}}},
	\end{aligned}
\end{equation}
where now
\begin{equation}
	\label{eq:BC:defNDB}
	\widehat{\mathcal{N}}_2 := 4\widehat{\lambda}_{12}\widehat{\lambda}_{21} - \left[\widehat{\lambda}_{11} + 1 \right]\left[\widehat{\lambda}_{22} - 1\right].
\end{equation}

It is important to note that the $C_{ij}$ integration constants are fundamentally different, as they are determined by different asymptotic boundary conditions and correspond to different physics. In section \ref{section:scattering} we will see exactly how they relate to the physical cross-sections, but having defined them all separately is already important at the level of renormalizing the point-particle couplings, which we do next.

\subsection{Renormalization-Group Flows and Invariants}
\label{section:BCs:RG}

Next we move on to teasing out of the boundary condition \eqref{eq:BC:SmallRBC} exactly how the couplings $h_{ij}$ must be renormalized with $\epsilon$ to keep physical quantities independent of the regulator. One way to do so would be to differentiate \eqref{eq:BC:IC:solveCsLOA} and \eqref{eq:BC:IC:solveCsLOB} with respect to $\epsilon$ while holding the (physical) integration constants fixed, and solve the resulting differential equations. This approach turns out to be very difficult however, since the equations are highly coupled and tough to disentangle. Notice however that it is important to have knowledge of both $1\to X$ and $2\to X$ scattering to solve for all the elements of $\bf h$. This is not a coincidence. A unitary system requires a real action, which is enforced by the condition that $\bf h$ is Hermitian. At the same time, a unitary S-matrix for an $N$-species system has $N^2$ real degrees of freedom, which is the same dimension as an $N\times N$ Hermitian matrix. Consequently, connecting the point-particle couplings to physical quantities requires knowledge of the entire S-matrix, and so in our case must involve both $1\to X$ scattering and $2\to X$ scattering. Lastly, one final simplification can be made by observing that the phase of $h_{12}$ can be removed by a field redefinition, so for the special case of a two-species system we only have to deal with a real matrix of point-particle couplings.

The easiest approach to solving for the flows of the couplings $h_{ij}$ is to go back to the boundary conditions \eqref{eq:BC:IC:rbcA} and \eqref{eq:BC:IC:rbcB} for both $1\to X$ and $2\to X$ systems and solve directly for the individual elements of $\bf h$. This inversion is done in detail in appendix \ref{appendix:solvePPs}, and using the small-$r$ form \eqref{eq:BC:IC:bulkModesLO}, the point-particle couplings must take the following forms as functions of the regulator $\epsilon$.
\begin{subequations}
	\label{eq:BC:RG:running}
	\begin{align}
		\widehat \lambda_{11} &=  \frac{(1 - C_{11}(2ik_1\epsilon)^{-\zeta_1})(1 + C_{22}(2ik_2\epsilon)^{-\zeta_2}) + C_{21}C_{12}(2ik_1\epsilon)^{-\zeta_1}  (2ik_2\epsilon)^{-\zeta_2} }{(1 + C_{11}(2ik_1\epsilon)^{-\zeta_1})(1 + C_{22}(2ik_2\epsilon)^{-\zeta_2}) - C_{21}C_{12}(2ik_1\epsilon)^{-\zeta_1}(2ik_2\epsilon)^{-\zeta_2}}, \label{eq:BC:RG:running:a}\\[1em]
		\widehat \lambda_{12} &= \sqrt{\frac{m_2\zeta_1}{m_1\zeta_2}} \frac{C_{21}\left(\frac{k_1}{k_2}\right)^{-1/2}(2ik_1\epsilon)^{-\zeta_1/2}(2ik_2\epsilon)^{-\zeta_2/2} } {(1 + C_{11}(2ik_1\epsilon)^{-\zeta_1})(1 + C_{22}(2ik_2\epsilon)^{-\zeta_2}) - C_{21}C_{12} (2ik_1\epsilon)^{-\zeta_1}(2ik_2\epsilon)^{-\zeta_2} }, \label{eq:BC:RG:running:b} \\[1.3em]
		\widehat \lambda_{21} &= \sqrt{\frac{m_1\zeta_2}{m_2\zeta_1}}\frac{C_{12}\left(\frac{k_2}{k_1}\right)^{-1/2}(2ik_1\epsilon)^{-\zeta_1/2}(2ik_2\epsilon)^{-\zeta_2/2} } {(1 + C_{11}(2ik_1\epsilon)^{-\zeta_1})(1 + C_{22}(2ik_2\epsilon)^{-\zeta_2}) - C_{21}C_{12} (2ik_1\epsilon)^{-\zeta_1}(2ik_2\epsilon)^{-\zeta_2} }, \label{eq:BC:RG:running:c} \\[1.3em]
		\widehat \lambda_{22} &=  \frac{(1 + C_{11}(2ik_1\epsilon)^{-\zeta_1})(1 - C_{22}(2ik_2\epsilon)^{-\zeta_2}) +C_{21}C_{12} (2ik_1\epsilon)^{-\zeta_1}(2ik_2\epsilon)^{-\zeta_2} }{(1 + C_{11}(2ik_1\epsilon)^{-\zeta_1})(1 + C_{22}(2ik_2\epsilon)^{-\zeta_2}) - C_{21}C_{12} (2ik_1\epsilon)^{-\zeta_1}(2ik_2\epsilon)^{-\zeta_2} }, \label{eq:BC:RG:running:d}
	\end{align}
\end{subequations}
using the definitions \eqref{eq:BC:defHats}.

Equations \eqref{eq:BC:RG:running} (together with \eqref{eq:BC:IC:solveCsLOA}, \eqref{eq:BC:IC:solveCsLOB}, \eqref{eq:review:defEps}, and past work \cite{burgess_reduced_2018}) suggests the integration constants can each be characterized by a unique RG-invariant length-scale. To see how this works, define the scales $\epsilon_1,\, \epsilon_2$ and $\epsilon_3$ by the following relations:
\begin{equation}
	\label{eq:BC:RG:defEps}
	\begin{aligned}
		&C_{11} = -y_1(2ik_1\epsilon_1)^{\zeta_1}, \qquad C_{22} = -y_2(2ik_2\epsilon_2)^{\zeta_2}, \\[1em]
		\text{and}\quad &C_{12} = \frac{m_2k_2\zeta_1}{m_1k_1\zeta_2}C_{21} = y_3\sqrt{\frac{m_2k_2\zeta_1}{m_1k_1\zeta_2}}(2ik_1\epsilon_3)^{\zeta_1/2}(2ik_2\epsilon_3)^{\zeta_2/2},
	\end{aligned}
\end{equation}
where $y_i = \pm 1$ defines a particular class of flow. In terms of these scales, the running equations are significantly simpler:
\begin{subequations}
	\label{eq:BC:RG:runningInv}
	\begin{align}
		\widehat \lambda_{11} &=  \frac{\left(1 + y_1\left( {\epsilon}/{\epsilon_1} \right) ^{-\zeta_1}\right)\left(1 - y_2\left( {\epsilon}/{\epsilon_2} \right) ^{-\zeta_2}\right) + \left( {\epsilon}/{\epsilon_3} \right)^{-(\zeta_1 + \zeta_2)} }{\left(1 - y_1\left( {\epsilon}/{\epsilon_1} \right) ^{-\zeta_1}\right)\left(1 - y_2\left( {\epsilon}/{\epsilon_2} \right) ^{-\zeta_2}\right) - \left( {\epsilon}/{\epsilon_3} \right)^{-(\zeta_1 + \zeta_2)} }, \label{eq:BC:RG:runningInv:a}\\[1em]
		\widehat \lambda_{12} &= \widehat \lambda_{21} = \frac{y_3\left( {\epsilon}/{\epsilon_3} \right)^{-(\zeta_1+ \zeta_2)/2} }{\left(1 - y_1\left( {\epsilon}/{\epsilon_1} \right) ^{-\zeta_1}\right)\left(1 - y_2\left( {\epsilon}/{\epsilon_2} \right) ^{-\zeta_2}\right) - \left( {\epsilon}/{\epsilon_3} \right)^{-(\zeta_1 + \zeta_2)} }, \label{eq:BC:RG:runningInv:b}\\[1em]
		\widehat \lambda_{22} &=  \frac{\left(1 - y_1\left( {\epsilon}/{\epsilon_1} \right) ^{-\zeta_1}\right)\left(1 + y_2\left( {\epsilon}/{\epsilon_2} \right) ^{-\zeta_1}\right) + \left( {\epsilon}/{\epsilon_3} \right)^{-(\zeta_1 + \zeta_2)} }{\left(1 - y_1\left( {\epsilon}/{\epsilon_1} \right) ^{-\zeta_1}\right)\left(1 - y_2\left( {\epsilon}/{\epsilon_2} \right) ^{-\zeta_2}\right) - \left( {\epsilon}/{\epsilon_3} \right)^{-(\zeta_1 + \zeta_2)} }, \label{eq:BC:RG:runningInv:c}
	\end{align}
\end{subequations}

\begin{figure}[ht]
		\centering
		\includegraphics[width=0.95\linewidth]{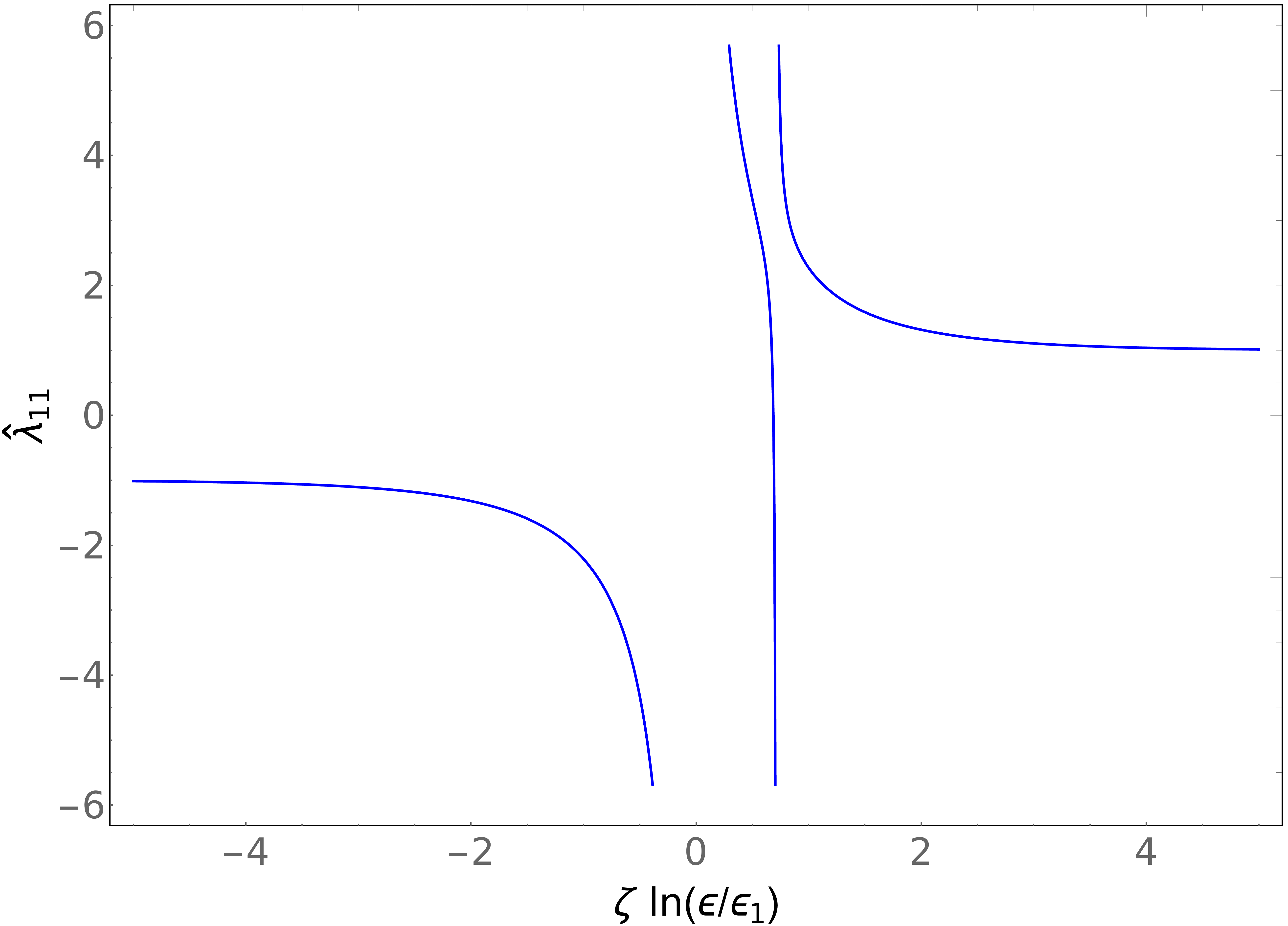}
		\caption{Plot of $\widehat \lambda_{11}$ vs $\zeta\ln(\epsilon/\epsilon_1)$ for $\zeta_1 = \zeta_2$, $y_1 = y_2 = +1$, and $(\epsilon_2/\epsilon_1)^\zeta = 2$ and $(\epsilon_3/\epsilon_1)^\zeta = 0.02$. The fascinating second pole arises in certain limits of the ratio of $\epsilon_1$ to $\epsilon_2$ and $\epsilon_3$. In the limit $\epsilon_3 \to 0$, this reduces to the classic single-particle RG.}
		\label{fig:plotM11}
	\end{figure}

\begin{figure}[ht]
	\centering
	\includegraphics[width=0.95\linewidth]{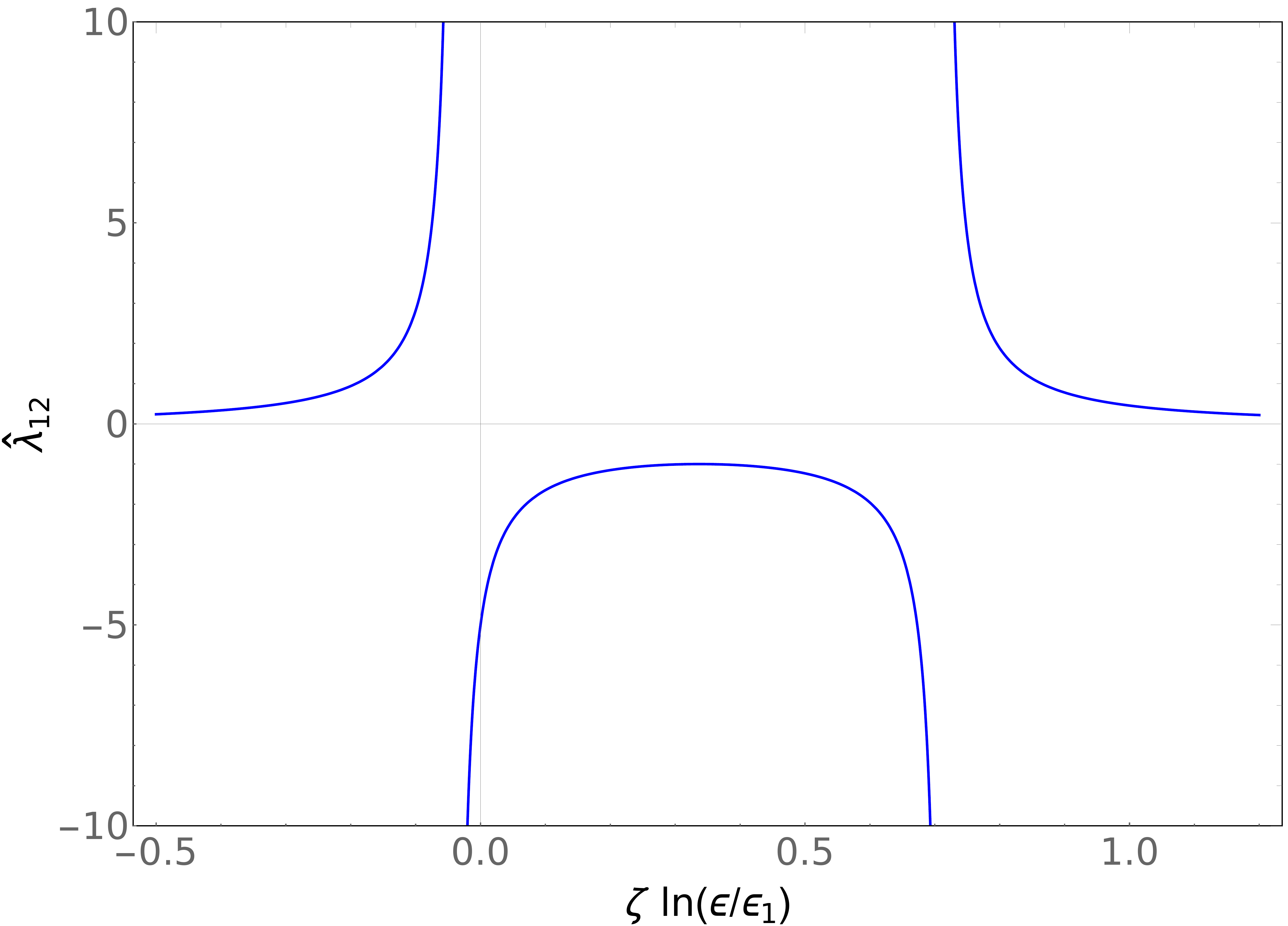}
	\caption{Plot of $\widehat \lambda_{12}$ vs $\zeta\ln(\epsilon/\epsilon_1)$ for $\zeta_1 = \zeta_2$, $y_1 = y_2 = y_3 = +1$, and $(\epsilon_2/\epsilon_1)^\zeta = 2$ and $(\epsilon_3/\epsilon_1)^\zeta = 0.02$. }
	\label{fig:plotM12}
\end{figure}

An example of $\widehat\lambda_{11}$ and $\widehat \lambda_{22}$ is plotted in figure \ref{fig:plotM11}, and an example of $\widehat \lambda_{12}$ is plotted in figure \ref{fig:plotM12}. All the couplings flow to fixed points in the ultraviolet ($\epsilon/\epsilon_i \to 0$) and the infrared ($\epsilon/\epsilon_i \to \infty$). The diagonal couplings $\widehat \lambda_{11}$ and $\widehat \lambda_{22}$ both flow to $-1$ in the UV and $+1$ in the IR, exactly as the single-particle flow does, while the off-diagonal $\widehat \lambda_{12}$ flows to vanishing coupling in both the cases. This says something reasonable: the system is perfectly content to live in a world where there is no species mixing, regardless of the existence of the inverse-square potential. However, if the species do mix, then the strength of that mixing depends on the scale it is measured at, with the flow given by \eqref{eq:BC:RG:runningInv:b}.

	Interestingly, all three flows share a common denominator, which can always be factorized. When $\zeta_1 = \zeta_2 =: \zeta$, the zeroes of the denominator lie at 
	\begin{equation}
		\label{eq:BC:RG:asym}
			\epsilon_a^{\zeta} = \frac{1}{2}\left(  y_2 \epsilon_2^{\zeta} + y_1 \epsilon_1^{\zeta} \pm \sqrt{\left(y_1\epsilon_1^{\zeta} - y_2\epsilon_2^{\zeta} \right)^2 + 4\epsilon_3^{2\zeta}}  \right),
	\end{equation}
	so there is at least one asymptote in all the flows as long as $y_2 \epsilon_2^{\zeta} + y_1 \epsilon_1^{\zeta} > 0$. Indeed, the only regime where there is \emph{no} asymptote is where $-(y_2 \epsilon_2^{\zeta} + y_1 \epsilon_1^{\zeta}) > \sqrt{\left(y_1\epsilon_1^{\zeta} - y_2\epsilon_2^{\zeta} \right)^2 + 4\epsilon_3^{2\zeta}} > 0$.

	The practicality of this framework lies in how the point-particle couplings (and in particular their RG-invariants \eqref{eq:BC:RG:defEps}) inform physical quantities, like scattering cross-sections, and we investigate this next.

\section{Scattering and Catalysis of Flavour Violation}
\label{section:scattering}

To see how the nuclear properties enter into macroscopic quantities, and in particular how the point-particle can induce a violation of bulk flavour-conservation, we proceed to compute the elements of the scattering matrix. In particular, it will be shown that the low-energy $s$-wave ``elastic'' ($\Psi_i \to \Psi_i$) scattering is as usual independent of  incoming momentum, and $\epsilon_i$ plays the role of the scattering length. Meanwhile, the flavour-violating ``inelastic'' cross-section is uniquely characterized by $\epsilon_3$, which can be thought of as an effective scattering length for flavour violation. Moreover, the inelastic cross-section goes as $k_\text{out}/k_\text{in}$. This ratio is a constant for Schr\"odinger particles, but for Klein-Gordon fields (such as is appropriate for, say, incident pions) the ratio has a dependence on the incoming momentum, and that dependence takes on a variety of qualitatively different forms determined by the size of $k_\text{in}^2/\abs{m_1^2 - m_2^2}$. Of particular interest is the case where the incoming particle's mass is greater than the outgoing particle's mass, and the incoming particle's momentum is small compared to the mass gap, in which case the flavour-violating cross-section is catalysed, and goes as $1/k_\text{in}$ (although such a setup would certainly be hard to realize in practice).

Following the appendix \ref{appendix:scattering:Multi}, we need to identify the scattering matrix elements $S_{ij}^{(\ell)}$ in terms of $C_{ij}$. 

\subsection{Elastic Scattering (\texorpdfstring{$\psi_i \to \psi_i$}{i -> i})}
\label{section:scattering:11}

First consider elastic scattering of $\psi_1$. This is the case where the measured particle is the same as the incoming particle, so the asymptotic form of $\psi_1$ contains an incoming plane wave and an outgoing spherical wave:
\begin{equation}
	\label{eq:scattering:11:ansP1}
	\psi_1^{\text{Ans}}(r \to \infty) \to \mathcal{C}\qty(e^{ik_1z} + f_{11}(\theta, \psi) \frac{e^{ik_1r}}{r}).
\end{equation}

Starting from the general form \eqref{eq:action:bulkFuncs} for particle 1,
\begin{equation}
	\label{eq:scattering:11:genP1}
	\psi_1 = C_{1+} \psi_{1+} + C_{1-} \psi_{1-},
\end{equation}
the large-$r$ limit is (taking the asymptotic limit of the confluent hypergeometric function)
\begin{equation}
	\label{eq:scattering:11:asymP1}
	\psi_{1}(r\to \infty) \to A_\ell\frac{e^{i(k_1 r - \ell\pi/2)}}{2ik_1r} + B_\ell\frac{e^{-i(k_1 r - \ell\pi/2)}}{2ik_1r},
\end{equation}
where
\begin{equation}
	\label{eq:scattering:11:relABC}
	\begin{split}
		A_\ell &= \qty[\Gamma\left(1 + \frac 12 \zeta_1\right)2^{\zeta_1} C_{1+} + \Gamma\left(1 - \frac 12 \zeta_1\right)2^{-\zeta_1}C_{1-}]\frac{e^{i\pi\ell/2}}{\sqrt{\pi}}, \quad \text{and}\\
		B_\ell &= \qty[\Gamma\left(1 + \frac 12 \zeta_1\right)2^{\zeta_1} C_{1+} + \Gamma\left(1 - \frac 12 \zeta_1\right)2^{-\zeta_1}e^{-i\pi\zeta_1}C_{1-}]\frac{e^{i(1 + \zeta_1 - \ell)\pi/2}}{\sqrt{\pi}}.
	\end{split}
\end{equation}
Matching to the asymptotic form \eqref{eq:scattering:11:ansP1}, allows us to identify (see \ref{appendix:scattering:singleElastics}) the overall normalization
\begin{equation}
	\label{eq:scattering:11:matchC}
	\mathcal{C} = \frac{(-1)^{\ell + 1}}{2\pi\sqrt{(2\ell + 1)}}\qty[\Gamma\left(1 + \frac 12 \zeta_1\right)2^{\zeta_1} C_{1+} + \Gamma\left(1 - \frac 12 \zeta_1\right)2^{-\zeta_1}e^{-i\pi\zeta_1}C_{1-}]e^{i(1 + \zeta_1)\pi/2},
\end{equation}
and the scattering matrix element
\begin{align}
	\label{eq:scattering:11:matchF1}
	S_{11}^{(\ell)} = - \frac{A_\ell}{B_\ell} &= \frac{\qty[\Gamma\left(1 + \zeta_1/2\right) + \Gamma\left(1 - \zeta_1/2\right)2^{-2\zeta_1}C_{11}]}{\qty[\Gamma\left(1 + \zeta_1/2\right) + \Gamma\left(1 - \zeta_1/2\right)2^{-2\zeta_1}e^{-i\pi\zeta_1}C_{11}]} e^{i(2\ell + 1 - \zeta_1)\pi/2}. 
\end{align}

Our interest is in the regime where $k_i\epsilon$ is small, so the $s$-wave is the dominant contribution to the cross-section. In the $s$-wave, $\zeta_1 = \zeta_{1s} := \sqrt{1 - 8m_1g}$, and the cross-section is \eqref{eq:app:sc:ss:totalXs}, exactly as in \cite{burgess_point-particle_2017-1}:
\begin{equation}
	\label{eq:scattering:11:X0}
	\sigma_s^{(1\to 1)} = \frac{\pi}{k_1^2}\abs*{S_{11}^{(0)}}^2 = \frac{\pi}{k_1^2}\abs{\frac{1 - \mathcal{A}e^{i\pi\zeta_{1s}/2}}{1 - \mathcal{A}e^{-i\pi\zeta_{1s}/2}}}^2
\end{equation}
where
\begin{equation}
	\label{eq:scattering:11:defA}
	\mathcal{A} := e^{-i\pi\zeta_{1}/2}2^{-2\zeta_1}C_{11}\frac{\Gamma[1- \frac{1}{2}\zeta_1]}{\Gamma[1+ \frac{1}{2}\zeta_1]} = y_1 \left( \frac{k_1\epsilon_1}{2} \right)^{\zeta_1}  \frac{\Gamma[1- \frac{1}{2}\zeta_1]}{\Gamma[1+ \frac{1}{2}\zeta_1]} .
\end{equation}
(The second equality uses \eqref{eq:BC:RG:defEps} to exchange the integration constants for the RG-invariant $\epsilon_1$). Of particular note is when there is no inverse-square potential, in which case $\zeta_{1s} = 1$ and the cross-section reduces to
\begin{equation}
	\label{eq:scattering:11:X0Zeta1}
	\sigma_s^{(1 \to 1)} = 4\pi\epsilon_1^2 \qquad (g = 0),
\end{equation}
which can be identified as the cross-section for scattering from a 3D $\delta$-function potential (see for example \cite{jackiw_delta_1991} where our $\epsilon_1$ corresponds to their $g/\sqrt{\pi}$). Elastic scattering for the second species $\psi_2 \to \psi_2$ follows exactly the same procedure and is trivially the $1 \leftrightarrow 2$ inversion of \eqref{eq:scattering:11:X0} and \eqref{eq:scattering:11:defA}.
\begin{equation}
	\label{eq:scattering:22:X0Zeta1}
	\sigma_s^{(2 \to 2)} = 4\pi\epsilon_2^2 \qquad (g = 0).
\end{equation}
%

%

\subsection{Flavour-Violating Scattering (\texorpdfstring{$\psi_i \to \psi_j$, $i \neq j$}{1 -> 2})}
\label{section:scattering:12}

So much for the ordinary scattering. Finally we compute the flavour-violating $\psi_1 \to \psi_2$ cross-section, and see the point-particle catalysis in action. This time, there is no incoming flux of the particle to be measured, so the large-$r$ ansatz is simply:
\begin{equation}
	\label{eq:mpScatt:ansP2}
	\psi_2^{\text{Ans}}(r \to \infty) \to \mathcal{C} f_2(\theta, \psi) \frac{e^{ik_2r}}{r}.
\end{equation}
where we have scaled out the same normalization factor $\mathcal{C}$, for convenience. The asymptotic form of the general solution for $\psi_2$ is exactly \eqref{eq:scattering:11:asymP1} subject to $1 \leftrightarrow 2$, so we can immediately observe the following. First, as was used in deriving the boundary conditions in section \ref{section:BCs}, $B_\ell = 0$ so that $C_{2+} = R\,C_{2-}$, where as above
\begin{equation}
	\label{eq:scattering:12:defR}
	R := -\frac{\Gamma(1 - \zeta/2)}{\Gamma(1 + \zeta/2)}2^{-2\zeta}e^{-i\pi\zeta}.
\end{equation}
%
Following appendix \ref{appendix:scattering:Multi}, this time we identify the inelastic scattering element $S_{12}^{(\ell)}$ using \eqref{eq:app:sc:ms:f2l} 
\begin{align}
	\label{eq:scattering:12:f12}
	S_{12}^{(\ell)} &= \frac{e^{-i\pi \ell/2}}{\sqrt{4\pi(2\ell + 1)}}\frac{A_{\ell}}{\mathcal{C}} = \frac{\Gamma\left( 1-\frac{1}{2}\zeta_2 \right)}{2\pi\sqrt{(2\ell + 1)}}\frac{C_{2-}}{\mathcal{C}} 2^{-\zeta_2}(1-e^{-i\pi\zeta_2}), \notag \\
				&= \frac{(-1)^{-\ell + 1}}{2^{\zeta_1+\zeta_2 - 1}}\sin(\pi\zeta_2/2)e^{i\pi (\ell - \zeta_1 - \zeta_2 - 1)/2}\frac{\Gamma\left( 1-\frac{1}{2}\zeta_2 \right)}{\Gamma\left( 1+\frac{1}{2}\zeta_1 \right)} \frac{C_{12}}{1 - \mathcal{A}e^{-i\pi\zeta_{1}/2}}. 
\end{align}
where $\mathcal{A}$ is defined in \eqref{eq:scattering:11:defA}.

Again, our interest is in the small $k_i\epsilon$ regime for which the $s$-wave dominates. We similarly define $\zeta_{2s} := \sqrt{1 - 8m_2g}$. Then the low-energy cross-section is \eqref{eq:app:sc:ms:multiX1to2},  
\begin{align}
	\label{eq:scattering:12:X0}
	\sigma_s^{(1\to 2)} &= \frac{\pi}{k_1k_2}\frac{m_1}{m_2}\abs{S_{12}^{(0)}}^2, \notag \\
						&= \frac{4\pi}{2^{\zeta_1+\zeta_2}}\frac{k_1^{\zeta_1}k_2^{\zeta_2}}{k_1^2}\frac{\zeta_1}{\zeta_2}\sin[2]( \pi\zeta_2/2 ) \frac{\epsilon_3^{\zeta_1+\zeta_2}}{\abs{1 - \mathcal{A}e^{-i\pi\zeta_1/2}}^2}\left(\frac{\Gamma\left( 1-\frac{1}{2}\zeta_2 \right)}{\Gamma\left( 1+\frac{1}{2}\zeta_1 \right)}\right)^2.
\end{align}
As in the elastic case, the reverse scattering $\psi_2 \to \psi_1$ is a simple matter of exchanging $1 \leftrightarrow 2$ in \eqref{eq:scattering:12:X0}. In the absence of an inverse-square potential ($\zeta_{is} = 1$), the cross-section \eqref{eq:scattering:12:X0} simplifies significantly.
\begin{equation}
	\label{eq:scattering:12:X0Zeta1}
	\sigma_{s}^{1\to 2} = 4\pi \frac{k_2}{k_1} \epsilon_3^2 \qquad (g = 0).
\end{equation}
On its own, this is an interesting enough result. The flavour-violating cross-section is non-zero only when the point-particle has non-trivial flavour-violating properties, as are encoded in $\epsilon_3$. This is also the statement that flavour-violation only occurs if $h_{12} \neq 0$, since it is always true---regardless of the presence of an inverse-square potential---that $h_{12} = 0$ only when $\epsilon_3 = 0$\footnote{As noted in section \ref{section:BCs:RG}, this is in contrast to the behaviour of $h_{11}$ and $h_{22}$, whose flows indicate that $\epsilon_1,\epsilon_2 \to 0$ is only consistent with $h_{11},h_{22} \to 0$ when there is no inverse-square potential.}.

A particularly interesting aspect of \eqref{eq:scattering:12:X0Zeta1} is the factor of $k_2/k_1$. For Schr\"odinger particles, this is just a constant ($k_i = \sqrt{2m_i E} \implies k_2/k_1 = \sqrt{m_2/m_2}$). In section \ref{section:rel:multiNuc} we treat a Schr\"odinger particle interacting with a multi-state nucleus, in which case the final and initial momenta do differ non-trivially, but even at the level of two bulk species, more interesting dynamics can be seen just by treating the particles as relativistic Klein-Gordon fields. For relativistic fields, the cross-section takes the same general form as \eqref{eq:scattering:12:X0Zeta1}, except the momenta are relativistic: $k_i = \sqrt{\omega^2 - m_i^2}$, where $\omega$ is the energy of the system. Then $\omega^2 = k_i^2 + m_i^2$, so that
\begin{equation}
	\label{eq:scattering:12:ratio}
	\frac{k_2}{k_1} = \frac{\sqrt{k_1^2 + (m_1^2 - m_2^2)}}{k_1}.
\end{equation}
The cross-section therefore exhibits different qualitative behaviour depending on how the incoming momentum $k_1$ relates to the (squared) mass gap $m_1^2 - m_2^2$. This can be broadly classified by 4 different regimes depending on the sign of $m_1^2 - m_2^2$ and the size of the ratio $r := k_1^2/\abs{m_1^2 - m_2^2}$. 

\begin{figure}[ht]
			\centering
			\includegraphics[width=0.95\linewidth]{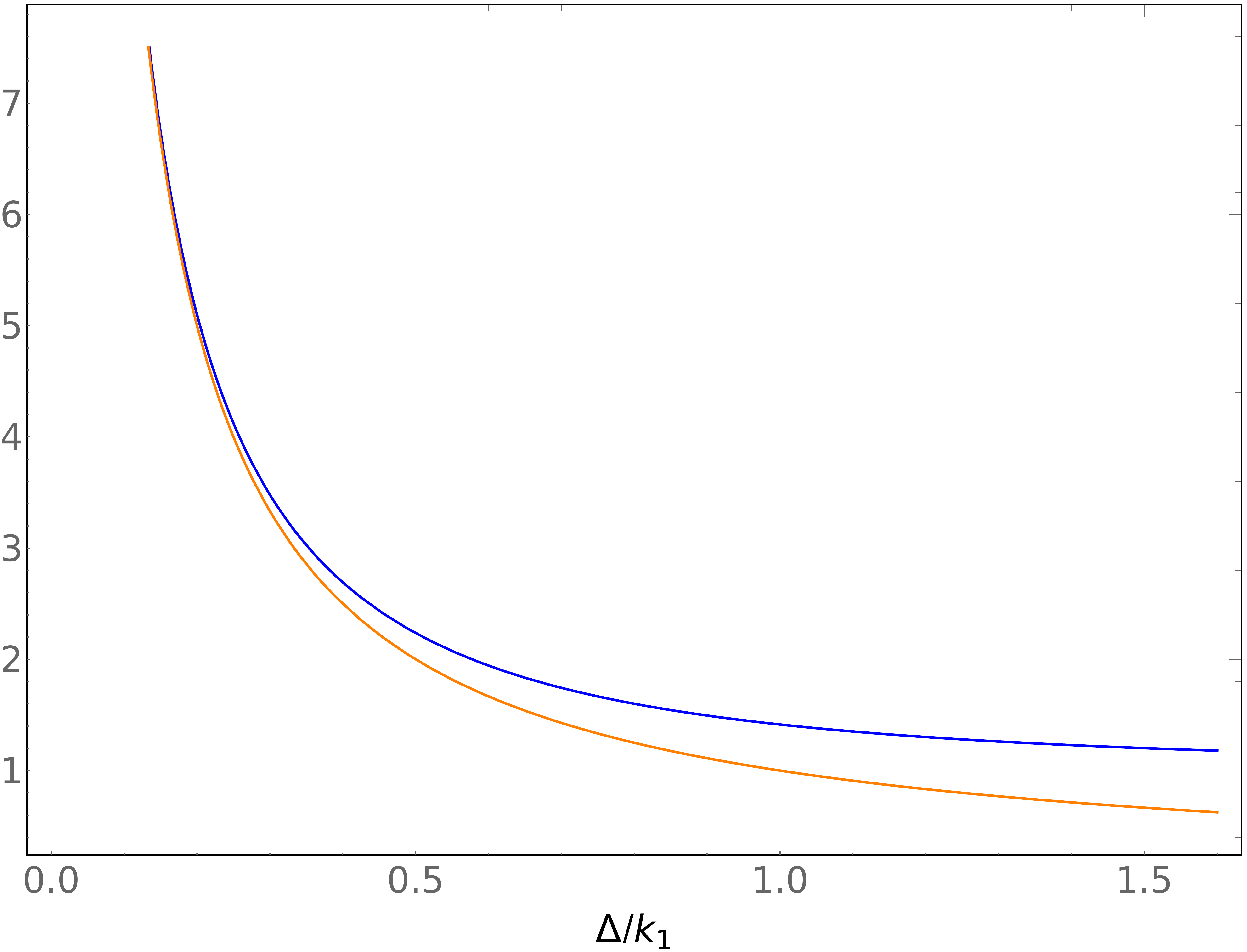}
			\caption{Plot of the cross-over behaviour in the $k$ dependence of the inelastic cross-section for a Klein-Gordon particle conversion in units of $\Delta/k_1$, where  $\Delta = \sqrt{m_1^2 - m_2^2}$ and $m_1 > m_2$. The full function $\sqrt{1+\Delta^2/k_1^2}$ is plotted in blue, and the simple inelastic behaviour $\Delta/k_1$ is plotted in orange. Notice the overlap for small $\Delta/k_1$, and the strong enhancement for large momenta.}
			\label{fig:crossover1}
		\end{figure}

First, if the mass gap is positive, the only question is to the size of the ratio $r$. If $r \ll 1$, then $\sigma_s^{(1\to 2)} \sim k_1^{-1}$. This is the regime where the cross-section sees a low-energy enhancement similar to the well-known enhancement of absorptive cross-sections \cite{bethe_theory_1935} (more on that in section \ref{section:rel:multiNuc}). However if $r \gg 1$, then the cross-section is roughly independent of the incoming momentum altogether. The cross-over between these regimes is plotted in figure \ref{fig:crossover1}. These are indeed reasonable behaviours. If $m_1 > m_2$, and the incoming momentum is small compared to the mass difference, then the transition is to a lighter particle travelling faster, which is intuitively a more favourable process---the heavier incident particle has access to a larger phase-space than the lighter incident particle. If the mass difference is small compared to the incoming momentum, then the benefit of transitioning to a particle with a smaller mass is minimal, so the process is no more favourable than no transition.

\begin{figure}[ht]
			\centering
			\includegraphics[width=0.95\linewidth]{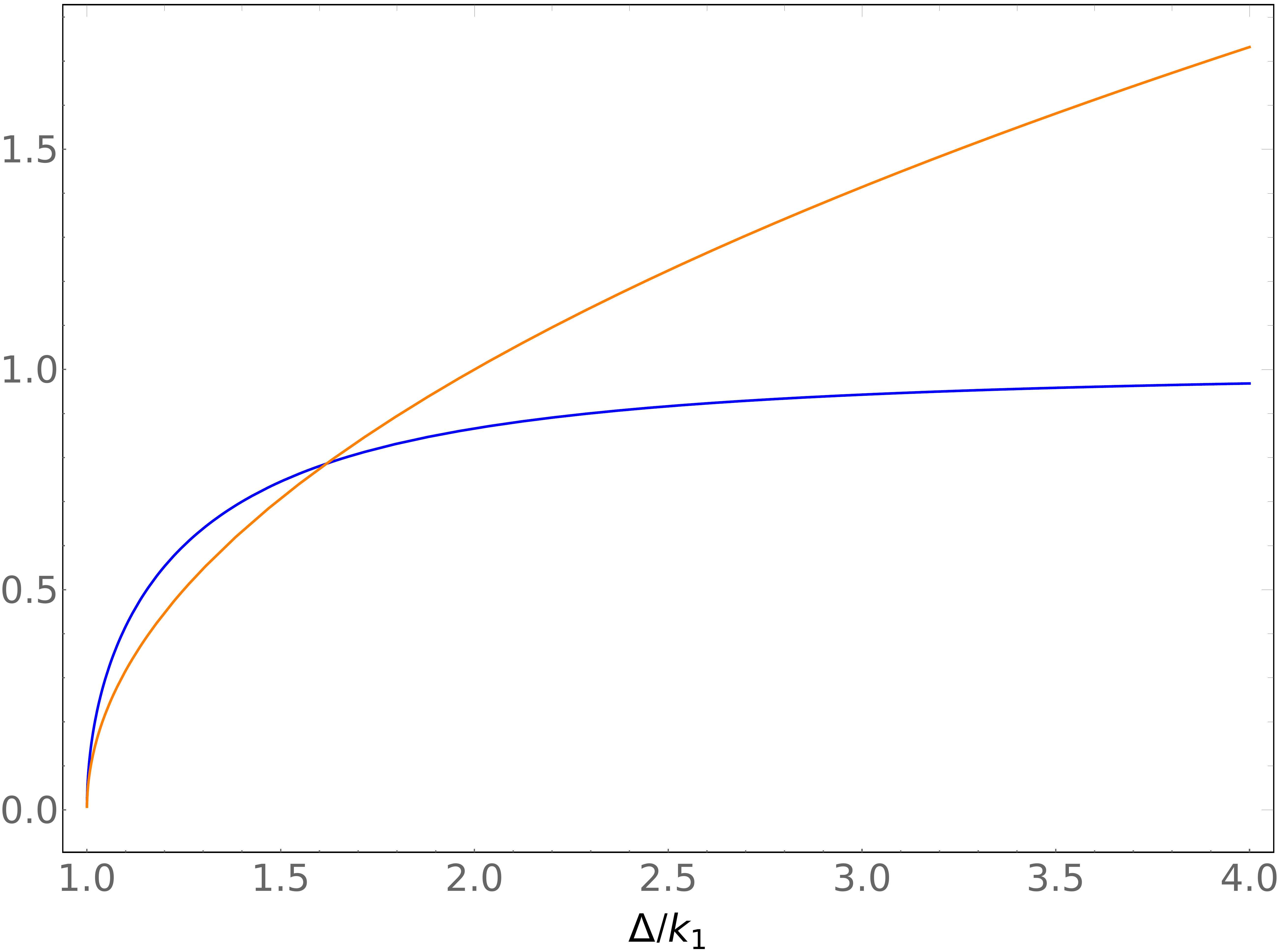}
			\caption{Plot of the cross-over behaviour in the $k$ dependence of the inelastic cross-section for a Klein-Gordon particle conversion in units of $\Delta/k_1$, where  $\Delta = \sqrt{m_1^2 - m_2^2}$ and $m_2 > m_1$. The full function $\sqrt{1-\Delta^2/k_1^2}$ is plotted in blue, and the simple inelastic behaviour $\sqrt{\Delta/k_1 - 1}$ is plotted in orange. Notice the threshold cutoff at $\Delta/k_1 = 1$, as well as the approximate overlap for small $\Delta/k_1$, and the strong suppression for large momenta.}
			\label{fig:crossover2}
		\end{figure}

		If instead the mass gap is negative, then there there is a new regime. If $k_1^2 < m_2^2 - m_1^2$  (so if $r < 1$) there is in fact \emph{no} scattering. This is certainly reasonable---if the incident particle did not have enough energy to create the rest mass of the second particle, then it cannot scatter into that particle (this is the threshold behaviour described by \cite[\S 144]{landau_chapter_1977}). If $r \gtrsim 1$, the incident momentum is just enough to create the second particle $k_1^2 = m_2^2 - m_1^2 + \delta$, then the cross-section goes as $\sqrt{\delta/(m_2^2 - m_1^2)}$. Since $k_2^2 = k_1^2 + m_1^2 - m_2^2 = \delta \ll m_2^2$, this is also the statement that the cross-section goes as $v_2$, the (non-relativistic) speed of the final state particle. Finally, if $r \gg 1$, the incident momentum greatly exceeds the mass gap and we again see the cross-section behave independently from $k_1$, as before. These momentum-dependences are plotted in figure \ref{fig:crossover2}.

\section{Transfer Reactions and Nuclear Structure}
\label{section:rel:multiNuc}

In many cases, a reaction with a nucleus can change not only the incident particle, but also the nucleus. This can be the case even when scattering energies are low compared to the mass of the nucleus. For instance, the excitation energy of most real nuclei is on the order of MeV compared to their masses of order GeV \cite{audi_nubase2016_2017}. A particularly interesting class of reactions that falls into this category is transfer reactions, where a composite particle (say a neutron) scatters off of a nucleus and exchanges a constituent particle (say a quark) with one of the valence nucleons, so that the outgoing particle is different (perhaps a proton) and so is the nucleus. While this work isn't enough to describe a complete transfer reaction, we can make progress towards a complete description, and can at least describe the simpler process $\psi + N \to  \psi + N^*$, where $N$ is some nucleus and $N^*$ is a long-lived excited state of that nucleus. The key observation to make is that there is essentially no difference between a system spanned by $\{\Psi_1\otimes\ket{N},\Psi_2\otimes\ket{N}\}$ with $\ket{N}$ some nuclear state, and $\{\Psi\otimes\ket{N_1},\Psi\otimes\ket{N_2}\}$ where $\ket{N_i}$ are distinct nuclear states. The only complexity lies in describing the different nuclear states in a point-particle EFT language.

Here we will sketch out the simplest point-particle action that includes a two-state nucleus coupled to a single bulk field, but a more detailed treatment of a PPEFT for a point-particle with internal degrees of freedom will be available in \cite{zalavari_ppeft_nodate}. In addition to the single-species action \eqref{eq:review:PPaction}, introduce an auxiliary grassmann-valued field $T_i$ that satisfies the commutation relations of the generators of $\mathfrak{su}(2)$.
\begin{equation}
	\label{eq:ME:brane}
	S^{\text{2N}}_{b} = -\int \dd \tau \sqrt{-\dot y^2}  \,\left(M + iT_i\dot T^i - i\epsilon_{ijk}\Delta_iT_jT_k  + h^\prime \abs{\Psi(y(\tau))}^2 - \epsilon_{ijk}g^\prime_i T_jT_k \abs{\Psi(y(\tau))}^2 \right).
\end{equation}
Here, $\Delta_i$ and $g_i^\prime$ are 3-vector-valued parameters. For convenience, we work in the basis such that $\Delta_i = \Delta\delta_{i3}$. Furthermore, we collect the point-particle couplings involving $\Psi$ as $W = W^\dagger := h^\prime - \frac{i}{2}g^\prime_iT_i$.

Upon quantization, the $T_i$ can be identified with $\frac{i}{2}\sigma_i$, the generators of $\mathfrak{su}(2)$, and since they only live on the point-particle's world-line, it is easy to see that they are associated with a two-level nuclear state. Varying the action with respect to $y(\tau)$ (and neglecting the subdominant contribution from the $W$ interactions), the nuclear dispersion relation 
\begin{equation}
	\label{eq:rel:2N}
	\mel{N}{\hat p_N^2 - m^2 + \Delta\sigma_z}{N} = 0, 	
\end{equation}
($\hat p_N$ is the nuclear 4-momentum operator) leads to distinct nuclear states $\ket{\uparrow}$ with rest-frame energy $E_\uparrow = M + \Delta/2$ and $\ket{\downarrow}$ with rest-frame energy $E_\downarrow = M - \Delta/2$.

The bulk action for the system is exactly the single-particle Schr\"odinger action \eqref{eq:review:bulkAction}, and so the solutions for $\Psi$ are precisely \eqref{eq:review:bulkSol} and \eqref{eq:review:modes}. However the boundary condition \eqref{eq:review:BC} becomes:
\begin{equation}
	\label{eq:rel:2N:BC}
	\mel{\uparrow\downarrow}{\lim_{\epsilon \to 0}\, 4\pi\epsilon^2 \partial_\epsilon \Psi}{\Xi} = \mel{\uparrow\downarrow}{\lim_{\epsilon \to 0}\, W \Psi(\epsilon)}{\Xi},
\end{equation}
where $\ket{\Xi}$ is an appropriate Fock state, $\ket{\uparrow\downarrow}$ are the Fock states consisting of just the nucleus in the state with rest-frame energy $E_{\uparrow\downarrow}$, and $\Psi$ is interpreted as an operator-valued field. Since energy can now be exchanged between the electron and the nucleus, the individual energies of the electron and the nucleus are no longer good quantum numbers, and a general single-electron Fock state must be a linear combination $\ket{\Xi} = \ket{\Psi_\uparrow}\ket{\uparrow} + \ket{\Psi_\downarrow}\ket{\downarrow}$, where $\ket{\Psi_{\uparrow\downarrow}}$ has energy $\omega_{\uparrow\downarrow}$ satisfying $\omega_\uparrow + \Delta/2 = \omega_\downarrow - \Delta/2$. Then in terms of the mode-functions $\Psi_{\uparrow\downarrow}(x) = e^{-i\omega_{\uparrow\downarrow}}\psi_{\uparrow\downarrow}$ (satisfying $\Psi\ket{\Psi_{\uparrow\downarrow}} = \Psi_{\uparrow\downarrow}(x)\ket{\Psi_{\uparrow\downarrow}}$), the boundary condition \eqref{eq:rel:2N:BC} is in components, 
\begin{equation}
	\begin{split}
		\label{eq:rel:2N:modeBC}
		4\pi\epsilon^2\partial_\epsilon\psi_\uparrow - W_{\uparrow\uparrow}\psi_\uparrow(\epsilon) - W_{\uparrow\downarrow}\psi_{\downarrow}(\epsilon)
 &= 0 \qquad \text{and} \\
		4\pi\epsilon^2\partial_\epsilon\psi_\downarrow - W_{\downarrow\downarrow}\psi_\downarrow(\epsilon) - W_{\downarrow\uparrow}\psi_{\uparrow}(\epsilon) &= 0,
	\end{split}
\end{equation}
where we identified the $T_i$ with $\frac{1}{2}\sigma_i$ so that $W = h^\prime \mathbb{I}_{2\times 2} + \frac{1}{4}g_i^\prime\sigma_i$. The boundary condition \eqref{eq:rel:2N:modeBC} is now exactly the boundary condition \eqref{eq:BC:SmallRBC} with $W \leftrightarrow \mathbf{h}$ and $\psi_{\uparrow\downarrow}\leftrightarrow \phi_{1,2}$. 

Finally, defining $E := \omega_\uparrow + \Delta/2 = \omega_\downarrow - \Delta/2$, we observe $k_\uparrow^2 = 2m\omega_\uparrow = 2m(E - \Delta/2)$ and $k_\downarrow^2 = 2m\omega_\downarrow = 2m(E + \Delta/2)$. Choosing $\psi_\uparrow \leftrightarrow \psi_1$ and $\psi_\downarrow \leftrightarrow \psi_2$, we may then identify $k_1 = k_\uparrow$ and  $k_2 = k_\downarrow$. At this point, a complete analogy with the two-species system is established, and the results are tabulated into a dictionary relating the two in table \ref{tab:dict2N}. A true transfer reaction is one for which the final state involves both a different species of bulk particle \emph{and} an altered state of the nucleus, so evidently would be equivalent to a model of 4 bulk species coupled to a single-state point-particle. As it stands however, this system is sufficient to describe, say, the low-energy behaviour of a neutron that knocks a nucleus into its first excited state.

\begin{table}[ht]
	\centering
	\begin{tabular}{c c c}
		Two-State Nucleus &  & Two-Species Bulk \\ \hline
		$\psi_\uparrow, \, \psi_\downarrow$ & $\longleftrightarrow$ & $\psi_1, \, \psi_2$\\[1em]
		$k_\uparrow,\, k_\downarrow$ & $\longleftrightarrow$ & $k_1,\, k_2$\\[1em]
		$W$	   & $\longleftrightarrow$ & $\bf h$
	\end{tabular}
	\caption{The dictionary that maps quantities in a two-bulk-species theory to quantities in a theory of a single bulk-species coupled to a point-particle with two accessible internal degrees of freedom.}
	\label{tab:dict2N}
\end{table}

\section{Single-Particle Subsector} 
\label{section:rel:abs}

In many cases it is overkill to track all of the possible final state products of a particular interaction. This is especially the case in nuclear physics, where a summation over many unobserved final states is the basis of the highly successful optical model \cite{feshbach_nuclear_2003,dickhoff_recent_2019}. The price one pays for the convenience of ignoring certain states is the loss of unitarity, and such a non-unitary point-particle EFT was the subject of \cite{plestid_fall_2018}. Here we can provide a very simple explicit example of how this non-unitarity can emerge in a subsector of a larger unitary theory. We achieve this correspondence by matching physical quantities, in a procedure that is significantly simpler than e.g.~tracing the partition function over the states involving $\Psi_2$ \cite{feldman_decoherence_1991,burgess_eft_2015}.

From reference \cite{plestid_fall_2018}, the key to a point-particle inducing a violation of unitarity is allowing the point-particle coupling to be complex ($h$ from section \ref{section:review}). In that case, the running of the coupling is the same as \eqref{eq:review:run}, except now the constant $-y \to e^{i\alpha_\star}$ is complex. That is:
\begin{equation}
	\label{eq:rel:abs:run}
	\hat\lambda_c = \frac{1 - e^{i\alpha_\star}(\epsilon/\epsilon_\star)^{-\zeta}}{1 + e^{i\alpha_\star}(\epsilon/\epsilon_\star)^{-\zeta}},
\end{equation}
where $\hat\lambda_c$ is the now complex coupling. Similarly, the integration constant is analogous to \eqref{eq:review:defEps}, 
\begin{equation}
	\label{eq:rel:abs:defEps}
	\frac{C_-}{C_+} = (2ik\epsilon)^{\zeta}\frac{1-\hat\lambda_c}{1 + \hat\lambda_c} = (2ik\epsilon_\star)^{\zeta}e^{i\alpha_\star}.
\end{equation}

At this point, we essentially have everything we need. The ratio of integration constants $\frac{C_-}{C_+}$ is directly related to the physical quantities in the single-particle problem, so choosing to track either $\Psi_1$ or $\Psi_2$ tells us to equate \eqref{eq:rel:abs:defEps} to $C_{11}$ or $C_{22}$ (respectively), and from that determine how the RG-invariants and couplings are related. The only obstruction at this point is that \eqref{eq:BC:RG:defEps} would at face value suggest that $\alpha_\star = n\pi$ and there is no absorptive scattering. The error here is that inelastic scattering is a sub-leading effect\footnote{The way to see this is through the catalysis cross-section, \eqref{eq:scattering:12:X0Zeta1}. Absorptive scattering generically scales as $a/k$ for some absorptive scattering length $a$. The derived cross-section \eqref{eq:scattering:12:X0Zeta1} identifies $a \sim (k_2\epsilon_3)\epsilon_3$ and so is generically a subdominant effect in the point-particle EFT regime.}, and to see it at the level of $C_{11}$, we would need to have computed that ratio of integration constants to sub-leading order in $k_i\epsilon$. This is not in itself a particularly challenging endeavour, and is done in the appendix \ref{appendix:2PICs}. The result is that to sub-leading order, one finds \eqref{eq:app:2PICs:C1NLOSimple} (choosing to track $\Psi_1$, tracking $\Psi_2$ follows trivially)
\begin{equation}
	\label{eq:rel:abs:redefEps}
	C_{11} = -(2ik_1\epsilon)^{\zeta_1}\frac{\widehat{\mathcal{N}}_1}{\widehat{\mathcal{D}}}\left( 1 - 4R\frac{\hat\lambda_{12}\hat\lambda_{21}}{\widehat{\mathcal{N}}_1\widehat{\mathcal{D}}}(2ik_2\epsilon)^{\zeta_2} \right) = -y_1(2ik_1\epsilon_1)^{\zeta_1}(1 + i\delta \alpha_1) .
\end{equation}
From the last equality, we use that $\delta\alpha_1 \ll 1$ to define $e^{i\alpha_1} \approx -y_1(1 + i\delta\alpha_1)$ such that $\alpha_1 := n\pi + \delta\alpha_1$, with $n$ an integer that satisfies $y_1 := -e^{in\pi}$. Since \eqref{eq:rel:abs:redefEps} is a perturbative expression, we may use the leading order (in $k_i\epsilon$) expressions for the couplings in $\delta \alpha_1$, and simply evaluate them at $\epsilon = \epsilon_1$. An alternative but more tedious approach would be to substitute the second equality in \eqref{eq:rel:abs:redefEps} into $\hat \lambda_{11}$ (taking the whole function to sub-leading order in $k_i\epsilon$ as in \eqref{eq:app:PPs:solveM11}) and solve for $\delta\alpha_1$ by demanding $\hat \lambda_{11}$ remain real at sub-leading order, as it must. No matter the approach, the result is
\begin{equation}
	\label{eq:rel:abs:solveDelA}
	\delta\alpha_1 = -(2k_2\epsilon_3)^{\zeta_1} \left( \frac{\epsilon_3}{\epsilon_1} \right)^{\zeta_2} \frac{\abs{R}}{y_1} \sin(\pi\zeta_2/2).
\end{equation}

In this way, we have solved for the RG-invariant quantities (and so too the physical quantities) in the single-particle absorptive model in terms of the RG-invariants in the unitary two-species model simply by equating $C_{11}$ to $C_-/C_+$. In fact, we can do even better than that. We can determine how the coupling $\hat\lambda_c$ relates to the various $\widehat \lambda_{ij}$ couplings. To do so, we simply arrange for $\widehat {\mathcal{N}}_A/\widehat {\mathcal{D}}$ to take the form of $(\hat\lambda_c - 1)/(\hat \lambda_c + 1)$.
\begin{equation}
	\label{eq:rel:abs:findHatL}
	\frac{\widehat {\mathcal{N}}_1}{\widehat {\mathcal{D}}} = \frac{4\abs{\widehat{\lambda}_{12}}^2 - \left[\widehat{\lambda}_{11} - 1 \right]\left[\widehat{\lambda}_{22} + 1\right]}{4\abs{\widehat{\lambda}_{12}}^2 - \left[\widehat{\lambda}_{11} + 1 \right]\left[\widehat{\lambda}_{22} + 1\right]} = \frac{\frac{\widehat{\lambda}_{11}[\widehat \lambda_{22} + 1] - 4\abs{\widehat{\lambda}_{12}}^2}{[\widehat{\lambda}_{22} + 1]} - 1}{\frac{\widehat{\lambda}_{11}[\widehat \lambda_{22} + 1] - 4\abs{\widehat{\lambda}_{12}}^2}{[\widehat{\lambda}_{22} + 1]} + 1}.
\end{equation}
Evidently $\hat\lambda_c = \widehat{\lambda}_{11} - 4\abs*{\widehat \lambda_{12}}^2/[\widehat \lambda_{22} + 1]$. A check on this is to directly compute the combination $\widehat{\lambda}_{11} - 4\abs*{\widehat \lambda_{12}}^2/[\widehat \lambda_{22} + 1]$, in which case one finds it is exactly \eqref{eq:rel:abs:run} with $\epsilon_\star$ and $\alpha_\star$ defined as above. This dictionary between these models is laid out in table \ref{tab:dictAbs}.

\begin{table}[ht]
	\centering
	\begin{tabular}{c c c}
		Absorptive Single-Species &  & Unitary Two-Species \\ \hline
		$\epsilon_\star$ & $\longleftrightarrow$ & $\epsilon_1$\\[1em]
		$\alpha_\star$ & $\longleftrightarrow$ & $n\pi -(2k_2\epsilon_3)^{\zeta_1} \left( \frac{\epsilon_3}{\epsilon_1} \right)^{\zeta_2} \frac{\abs{R}}{y_1} \sin(\pi\zeta_2/2) $\\[1em]
		$\hat\lambda_c$	   & $\longleftrightarrow$ & $\widehat{\lambda}_{11} - 4\frac{\abs*{\widehat \lambda_{12}}^2}{[\widehat \lambda_{22} + 1]}$
	\end{tabular}
	\caption{The dictionary that maps quantities in a unitary two-species theory to quantities in a non-unitary single-species theory.}
	\label{tab:dictAbs}
\end{table}

\begin{acknowledgments}
	We thank L\'aszl\'o Zalav\'ari, Daniel Ruiz, Markus Rummel, and Ryan Plestid for helpful discussions.
	
	This research was supported in part by funds from the Natural Sciences and Engineering Research Council (NSERC) of Canada. Research at the Perimeter Institute is supported in part by the Government of Canada through Industry Canada, and by the Province of Ontario through the Ministry of Research and Information (MRI).  
\end{acknowledgments}
\appendix

\section{Multi-Particle Partial-Wave Scattering}
\label{appendix:scattering}

Here we review the general framework of partial-wave scattering, including discussion of inelastic and multi-channel scattering.

\subsection{Single Particle Elastic Scattering}
\label{appendix:scattering:singleElastics}

We consider a spinless particle scattering off of a spinless, infinitely massive object at rest at the origin, through a spherically symmetric interaction $V(r)$. As usual, we employ the ansatz that at large distances from the target, the wavefunction of the incident particle is the sum of a plane wave incident along the $z$-axis and a scattered spherical wave:
\begin{equation}
	\label{eq:app:sc:ss:AsymptPsiAns}
	\psi^{\text{Ans}}_{\infty}(r) \to \mathcal{C}\qty(e^{ikz} + f(\theta, \phi) \frac{e^{ikr}}{r}).
\end{equation}
The differential cross-section is the ratio of the flux of the scattered particles $F_{sc}$ to the flux of the incoming particles $F_{in}$. With an incident beam of $N$ particles, the incoming flux is $N\, \mathbf{j}_{in}\cdot \mathbf{e}_z = N\,\abs{\mathcal{C}}^2\,k/m$, and the scattered flux is $N \abs{\mathcal{C}}^2 \abs{f(\theta,\phi)}^2 k / m r^2$, so that and is given by
\begin{equation}
	\label{eq:app:sc:ss:defDiffX}
	\dv{\sigma}{\Omega} := \frac{F_{sc}}{F_{in}} = \frac{1}{F_{in}}\, \mathbf{j}_{sc}\cdot \mathbf{e}_r\, r^2 = \abs{f(\theta)}^2.
\end{equation}
(And for a spherically symmetric scatterer, $f(\theta,\phi) = f(\theta)$).

At the same time, we consider solutions to the full Schrodinger equation
\begin{equation}
	\label{eq:app:sc:ss:Schrodinger}
	\frac{1}{r^2}\partial_r\qty(r^2\partial_r\psi(r)) - \qty[\frac{\ell(\ell + 1)}{r^2} + 2mV(r) - k^2]\psi(r) = 0,
\end{equation}
with $k^2 := 2mE$, and the full wavefunction expanded in a series of spherical harmonics $\Psi(\vec{x}, t) = e^{-iEt}\psi(r)Y_{\ell 0}$ (where we set $m = 0$ due to conservation of angular momentum). The asymptotic form of these radial functions is:
\begin{equation}
	\label{eq:app:sc:ss:AsymptPsiSch}
	\psi^{\text{Sch}}_\infty(r) \to A_\ell \frac{e^{i(kr - \ell\pi/2)}}{2ikr} + B_\ell \frac{e^{-i(kr - \ell\pi/2)}}{2ikr},
\end{equation}

Finding $f(\theta)$ now amounts to matching \eqref{eq:app:sc:ss:AsymptPsiAns} and \eqref{eq:app:sc:ss:AsymptPsiSch}. This can be accomplished by writing the plane-wave $e^{ikz}$ in terms of Legendre polynomials. The standard expansion is given as \cite{landau_chapter_1977-1}
\begin{equation}
	\label{eq:app:sc:ss:expPlaneWave}
	e^{ikz} = \sum_{\ell = 0}^{\infty} (2\ell + 1)i^\ell j_\ell(kr) P_\ell(\cos(\theta)) \to \sum_{\ell = 0}^{\infty} (2\ell + 1)i^\ell \frac{e^{i(kr - \ell\pi/2)} - e^{-i(kr - \ell\pi/2)}}{2ikr} P_\ell(\cos(\theta)) .
\end{equation}
Choosing the Condon-Shortly phase convention, the spherical harmonics can be written\\
$Y_\ell^0 = \sqrt{(2\ell + 1)/4\pi}P_\ell(\cos(\theta))$, so by computing the difference:
\begin{equation}
	\label{eq:app:sc:ss:ScattDiff}
	\psi^{\text{Sch}}_\infty(r) - \mathcal{C} e^{ikz} = \mathcal{C} f(\theta)\frac{e^{ikr}}{r},
\end{equation}
one finds first:
\begin{equation}
	\label{eq:app:sc:ss:firstCondition}
	B_\ell = -\sqrt{4\pi(2\ell + 1)}i^\ell \mathcal{C},
\end{equation}
set by the fact that there can be no incoming wave in the scattered wavefunction. Finally, one finds 
\begin{align}
	\label{eq:app:sc:ss:secondSol}
	f(\theta) &= \frac{1}{2ik}\sum_{\ell = 0}^{\infty} (2\ell + 1)\qty[S_\ell - 1]P_\ell(\theta),
\end{align}
Where $S_\ell = -A_\ell/B_\ell$ is the scattering matrix element. When the scattering is elastic as we've just described ($k_{\text{out}} = k_\text{in}$), the matrix element $S_\ell = e^{2i\delta_\ell}$ is a pure phase. Otherwise, when the scattering is inelastic and probability is lost, $S_\ell$ is just a complex number, but it is still common \cite{feshbach_nuclear_2003} to parameterize it as $S_\ell^{(in)} = e^{2i\gamma_\ell}$, where $\gamma_\ell$ is now a complex number. 

Finally, the total cross-section is computed as the integral over the differential cross-section, which is (using the orthogonality of the Legendre polynomials)
\begin{equation}
	\label{eq:app:sc:ss:totalXs}
	\sigma = \int \dd\Omega \dv[]{\sigma}{\Omega} = \frac{\pi}{k^2}\sum_{\ell} (2\ell + 1) \abs{S_\ell - 1}^2.
\end{equation}

\subsection{Multi-Channel Scattering}
\label{appendix:scattering:Multi}

It is a simple matter to generalize the above to multi-channel scattering. We treat the case of two species of particles, but as detailed in section \ref{section:rel:multiNuc} the results are more general. Without loss of generality, we will only look at $1\to X$ scattering.

Following \cite{landau_chapter_1977}, we begin by assuming the asymptotic forms for each species:
\begin{equation}
	\label{eq:app:sc:ms:AsymptPsiAnsP1}
	\psi_1^{\text{Ans}}(r \to \infty) \to \mathcal{C}\qty(e^{ik_1z} + f_1(\theta, \phi) \frac{e^{ik_1r}}{r}).
\end{equation}
and
\begin{equation}
	\label{eq:app:sc:ms:AsymptPsiAnsP2}
	\psi_2^{\text{Ans}}(r \to \infty) \to \mathcal{C} f_2(\theta, \phi) \frac{e^{ik_2r}}{r}.
\end{equation}
The differential cross-sections are defined in exactly the same way as the above. This means the $1\to 1$ scattering is exactly given by \eqref{eq:app:sc:ss:totalXs}, while for $1\to 2$ scattering we have
\begin{equation}
	\label{eq:app:sc:ms:difXs2}
	\dv[]{\sigma^{1\to 2}}{\Omega} = \frac{k_2}{k_1}\frac{m_1}{m_2}\abs{f_2(\theta)}^2.
\end{equation}
Particle 2 satisfies the same Schr\"odinger equation \eqref{eq:app:sc:ss:Schrodinger}, and so has the same asymptotic form \eqref{eq:app:sc:ss:AsymptPsiSch}. Matching to the ansatz is then as simple as
\begin{equation}
	\label{eq:app:sc:ms:match}
	\psi_{2,\infty}^{\text{Sch}} = \mathcal{C}f_2(\theta)\frac{e^{ik_2r}}{r},
\end{equation}
which produces 
\begin{equation}
	\label{eq:app:sc:ms:B2l}
	B_{2\ell} = 0,	
\end{equation}
and
\begin{equation}
	\label{eq:app:sc:ms:f2l}
	f_{2}(\theta) = \frac{1}{2ik_2\mathcal{C}}\sum_\ell e^{-i\pi \ell/2}A_{2\ell}Y_{\ell}^{0}(\theta) = \frac{1}{2ik_2}\sum_\ell (2\ell + 1) S_{12}^{(\ell)} P_\ell(\cos\theta),
\end{equation}
which defines the scattering matrix element $S_{12}^{(\ell)} = e^{-i\pi \ell/2}[4\pi(2\ell + 1)]^{-1/2} A_{2\ell}/\mathcal{C}$. Finally, the total cross-section is
\begin{equation}
	\label{eq:app:sc:ms:multiX1to2}
	\sigma^{1\to2} = \frac{\pi}{k_1k_2}\frac{m_1}{m_2} \sum_\ell (2\ell + 1) \abs*{S_{12}^{(\ell)}}^2.
\end{equation}

One important observation: as long as $S_{12}^{(\ell)} \neq 0$ for any $\ell$, then $S_{11}^{(\ell)} = e^{2i\gamma_{11}^{(\ell)}}$ must satisfy that the phase $\gamma_{11}^{(\ell)}$ is complex, since some of the probability flux of the incident particle 1 must have transferred to particle 2. 

\section{Solving for 2-Particle Integration Constant Ratios}
\label{appendix:2PICs}

Here we outline the details of the main calculation in section \ref{section:BCs:IC}. We do this for $1\to X$ scattering, but the results are easily applied to $2\to X$ scattering by inverting $1\leftrightarrow 2$. 

We begin with the boundary condition \eqref{eq:BC:SmallRBC}, and the general forms
\begin{equation}
	\label{eq:app:2PICs:genPsi}
	\psi_i = C_{i+} \psi_{i+} + C_{i-}\psi_{i-},
\end{equation}
where $i = 1,2$. For $1 \to X$ scattering, we use $C_{2+} = R\, C_{2-}$ with $R$ defined in \eqref{eq:BC:IC:defR}, and the boundary conditions are:
\begin{subequations}
	\label{eq:app:2PICs:BC}
	\begin{align}
		\widehat\psi_{1+}^\prime + C_{11} \widehat \psi_{1-}^\prime &= h_{11}(\psi_{1+} + C_{11}\psi_{1-}) + h_{12}C_{12}(\psi_{2+}R + \psi_{2-}), \quad \text{and} \label{eq:app:2PICs:BC1} \\
		C_{12}(\widehat\psi_{2+}^\prime R + \widehat \psi_{2-}^\prime) &= h_{22}C_{12}(\psi_{2+}R_2 + \psi_{2-}) + h_{21}(\psi_{1+} + C_{11}\psi_{1-}). \label{eq:app:2PICs:BC2}
	\end{align}
\end{subequations}
where we define $\widehat \psi_{i\pm}^\prime := \frac{2\pi\epsilon^2}{m_i}\partial_\epsilon\psi_{i\pm}$, and as in \eqref{eq:BC:IC:defICA}, we define $C_{11} := C_{1-}/C_{1+}$ and $C_{12} := C_{2-}/C_{1+}$.

Rearranging \eqref{eq:app:2PICs:BC1} for $C_{12}$, one finds
\begin{equation}
	\label{eq:app:2PICs:solveC2First}
	C_{12} = \frac{\qty[\widehat \psi_{1+}^\prime - h_{11}\psi_{1+}] +  C_{11}\qty[\widehat \psi_{1-}^\prime - h_{11}\psi_{1-}]}{h_{12}\qty[\psi_{2+}R + \psi_{2-}]}.
\end{equation}
Substituting in \eqref{eq:app:2PICs:BC2}, 
\begin{equation}
	\label{eq:app:2PICs:firstSub}
	\qty{[\widehat\psi_{1+}^\prime - h_{11}\psi_{1+}] +  C_{11}[\widehat \psi_{1-}^\prime - h_{11}\psi_{1-}]}Z = \abs{h_{12}}^2[\psi_{1+} +  C_{11}\psi_{1-}],
\end{equation}
where
\begin{equation}
	\label{eq:app:2PICs:defZ}
	Z := \frac{\widehat\psi_{2+}^\prime R + \widehat\psi_{2-}^\prime}{\psi_{2+}R + \psi_{2-}} - h_{22}.
\end{equation}
Finally rearranging, we have
\begin{equation}
	\label{eq:app:2PICs:genC1}
	 C_{11} = -\frac{\psi_{1+}}{\psi_{1-}}\frac{\left[\abs{h_{12}}^2 - \left[\frac{\widehat\psi_{1+}^\prime}{\psi_{1+}} - h_{11}\right]Z\right]}{\left[\abs{h_{12}}^2 - \left[\frac{\widehat\psi_{1-}^\prime}{\psi_{1-}} - h_{11}\right]Z\right]}
\end{equation}
Plugging this back into \eqref{eq:app:2PICs:solveC2First}, we have
\begin{align}
	\label{eq:app:2PICs:genC2}
	 C_{12} &= \frac{\psi_{1+}}{\psi_{2-}}\frac{\qty[\frac{\widehat\psi_{1+}^\prime}{\psi_{1+}} - h_{11}]\qty[\abs{h_{12}}^2 - \left[\frac{\widehat\psi_{1-}^\prime}{\psi_{1-}} - h_{11}\right]Z] - \qty[+ \leftrightarrow -]}{h_{12}\qty[R\frac{\psi_{2+}}{\psi_{2-}} + 1]\qty[\abs{h_{12}}^2 - \left[\frac{\widehat\psi_{1-}^\prime}{\psi_{1-}} - h_{11}\right]Z]}, \notag \\
				  &= \frac{\psi_{1+}}{\psi_{2-}}\frac{\qty[\frac{\widehat\psi_{1+}^\prime}{\psi_{1+}} - \frac{\widehat\psi_{1-}^\prime}{\psi_{1-}}]h_{21}}{\qty[R\frac{\psi_{2+}}{\psi_{2-}} + 1]\qty[\abs{h_{12}}^2 - \left[\frac{\widehat\psi_{1-}^\prime}{\psi_{1-}} - h_{11}\right]Z]}.
\end{align}
The integration constants in the $2\to X$ system are solved for in the same way. Solutions for $C_{22} = C_{2-}/C_{2+}$ and $C_{21} = C_{1-}/C_{2+}$ in the $2\to X$ are obtained directly from \eqref{eq:app:2PICs:genC1} and \eqref{eq:app:2PICs:genC2} (respectively) by simply inverting $1 \leftrightarrow 2$. 

In order to make use of these formulae, we now have to take the small-$r$ limit of the mode functions to the appropriate order. 

\subsection{Leading-Order in \texorpdfstring{$k_i\epsilon$}{k epsilon}}

First, we make the usual leading-order approximation. For $\psi_1$, this is exactly as in \eqref{eq:BC:IC:bulkModesLO}: 
\begin{align}
	\label{eq:app:2PICs:smallrPsi1LO}
	\psi_1(\epsilon) &\approx x_1^{-1/2}\qty[C_{1+}x_1^{\zeta_1/2} + C_{1-}x_1^{-\zeta_1/2}], \quad \text{and} \notag \\
	\partial_r \psi_1(\epsilon) &\approx ik_1 x_1^{-3/2}\qty[(\zeta_1 - 1)C_{1+}x_1^{\zeta_1/2} - (\zeta_1 + 1)C_{1-}x_1^{-\zeta_1/2}],
\end{align}
where for convenience here we define $x_i := (2ik_i\epsilon)$, for $i = 1,2$. Here we keep both the divergent $(-)$ and the (often) finite $(+)$ term because the ratio $C_{1-}/C_{1+}$ arises from the point-particle dynamics, and so is of the order $k_i\epsilon \ll 1$, which allows the two terms in \eqref{eq:app:2PICs:smallrPsi1LO} to compete. For particle 2 however, this is not the case. $C_{2+} = R\, C_{2-}$ with $R \sim \order{1}$ so that there is no balancing of the modes, and the divergent mode is simply dominant. That is:
\begin{align}
	\label{eq:app:2PICs:smallrPsi2LO}
	\psi_2(\epsilon) &\approx C_{2-}x_2^{-1/2 - \zeta_2/2}, \quad \text{and} \notag \\
	\partial_r \psi_2(\epsilon) &\approx -(\zeta_2 + 1) ik_2 C_{2-} x_2^{-3/2 - \zeta_2/2}.
\end{align}
With these approximations, we compute:
\begin{align}
	\label{eq:app:2PICs:LOFactors}
	\frac{\widehat \psi_{i\pm}^\prime}{\psi_{i\pm}} &\approx -\frac{\pi\epsilon}{m_i} (1 \mp \zeta_i), \notag \\
	\frac{\psi_{1+}}{\psi_{1-}} &\approx (2ik_1\epsilon)^{\zeta_1}, \notag \\
	\frac{\psi_{2+}}{\psi_{2-}} &\approx 0, \notag \quad \text{and}\\
	Z &\approx \frac{\widehat \psi_{2-}^\prime}{\psi_{2-}} - h_{22} \approx -\qty[(1 + \zeta_2)\frac{\pi\epsilon}{m_2} + h_{22}].
\end{align}
Then to leading order in $k_i\epsilon$, it is found that:
\begin{equation}
	\label{eq:app:2PICs:C1LO}
	C_{11} \approx -(2ik_1\epsilon)^{\zeta_1}\frac{\left[\abs{h_{12}}^2 - \left[\frac{\pi\epsilon}{m_1} (1 - \zeta_1) + h_{11}\right]\qty[(1 + \zeta_2)\frac{\pi\epsilon}{m_2} + h_{22}]\right]}{\left[\abs{h_{12}}^2 - \left[\frac{\pi\epsilon}{m_1} (1 + \zeta_1) + h_{11}\right]\qty[(1 + \zeta_2)\frac{\pi\epsilon}{m_2} + h_{22}]\right]},
\end{equation}
and
\begin{equation}
	\label{eq:app:2PICs:C2LO}
	 C_{12} = (2ik_1\epsilon)^{\zeta_1/2}(2ik_2\epsilon)^{\zeta_2/2}\sqrt{\frac{k_2}{k_1}}\frac{2\pi\epsilon\zeta_1 h_{21}}{m_1\left[\abs{h_{12}}^2 - \left[\frac{\pi\epsilon}{m_1} (1 + \zeta_1) + h_{11}\right]\qty[(1 + \zeta_2)\frac{\pi\epsilon}{m_2} + h_{22}]\right]}
\end{equation}

It will become apparent that a redefinition of parameters can significantly clean up our equations. Drawing from the next appendix, we define
\begin{equation}
	\label{eq:app:2PICs:defHats}
	\begin{split}
		\hat\lambda_{11} := \frac{1}{\zeta_1}\left( \frac{m_1h_{11}}{\pi\epsilon} + 1 \right) &\qquad \hat\lambda_{12} := \frac{h_{12}\sqrt{m_1m_2} }{2\pi\epsilon\sqrt{\zeta_1\zeta_2} }, \\
		\hat\lambda_{22} := \frac{1}{\zeta_2}\left( \frac{m_2h_{22}}{\pi\epsilon} + 1 \right) &\qquad \hat\lambda_{21} := \frac{h_{21}\sqrt{m_1m_2} }{2\pi\epsilon\sqrt{\zeta_1\zeta_2} }.
	\end{split}
\end{equation}
In terms of these new variables, the integration constants read:
\begin{equation}
	\label{eq:app:2PICs:C1LOeps}
	C_{11} = -(2ik_1\epsilon)^{\zeta_1}\frac{\left[4\hat\lambda_{12}\hat\lambda_{21} - \left[\hat\lambda_{11} - 1\right]\qty[\hat\lambda_{22} + 1]\right]}{\left[4\hat\lambda_{12}\hat\lambda_{21} - \left[\hat\lambda_{11} + 1\right]\qty[\hat\lambda_{22} + 1]\right]},
\end{equation}
and
\begin{equation}
	\label{eq:app:2PICs:C2LOeps}
	 C_{12} = (2ik_1\epsilon)^{\zeta_1/2}(2ik_2\epsilon)^{\zeta_2/2}\sqrt{\frac{m_2k_2\zeta_1}{m_1k_1\zeta_2}}\frac{4\hat\lambda_{21}}{\left[4\hat\lambda_{12}\hat\lambda_{21} - \left[\hat\lambda_{11} + 1\right]\qty[\hat\lambda_{22} + 1]\right]}.
\end{equation}

\subsection{Sub-Leading-Order in \texorpdfstring{$k_i\epsilon$}{k epsilon}}

To leading order, $C_{11} \sim (2ik\epsilon)^{\zeta_1}$, so that to leading order $S_{11}^{(\ell)}$ is a pure phase. In order to see the emergence of an absorptive single-particle model in the particle 1 subsector of the theory (as covered in section \ref{section:rel:abs}), it is necessary to compute $C_{11}$ to the next order in $k_i\epsilon$. To that end, recall:
\begin{equation}
	\label{eq:app:2PICs:genPsiPM}
	\psi_\pm = (2ik\epsilon)^{\frac 12(-1 \pm \zeta)}e^{-ik\epsilon}\mathcal{M}\qty[\frac 12\qty(1 \pm \zeta), 1\pm\zeta, 2ik\epsilon],
\end{equation}
so that
\begin{align}
	\label{eq:app:2PICs:smallrPsiPMNLO}
	\psi_\pm &\approx (2ik\epsilon)^{\frac 12(-1 \pm \zeta)}\qty[1 - ik\epsilon + \order{(k\epsilon)^2}]\qty[1 + ik\epsilon + \order{(k\epsilon)^2}], \notag \\
			 &\approx (2ik\epsilon)^{\frac 12(-1 \pm \zeta)}\qty[ 1 + \order{(k\epsilon)^2}]
\end{align}
so at least for $\zeta < 2$, the leading correction is only in $\psi_2$, and is to include the $(+)$ mode, since it is only a factor of $(k_2\epsilon)^{\zeta_2}$ compared to the two powers from all other higher-order corrections. In fact this is a property only of systems without a $1/r$ potential, as in that case the leading correction from the hypergeometric factor does not cancel that from the exponential. 

Pushing through then, we repeat the calculation from the previous section, now using
\begin{align}
	\label{eq:app:2PICs:smallrPsi2NLO}
	\psi_2(\epsilon) &\approx C_{2-}x_2^{-1/2}\qty[x_2^{-\zeta_2/2} + R\, x_2^{\zeta_2/2}], \quad \text{and} \notag \\
	\partial_r \psi_2(\epsilon) &\approx  ik_2 C_{2-} x_2^{-3/2}\qty[-(\zeta_2 + 1)x_2^{-\zeta_2/2} + (\zeta_2 - 1)R\,x_2^{\zeta_2/2}].
\end{align}
With these approximations, we compute:
\begin{align}
	\label{eq:app:2PICs:NLOFactors}
	\frac{\hat \psi_{i\pm}^\prime}{\psi_{i\pm}} &\approx -\frac{\pi\epsilon}{m_i} (1 \mp \zeta_i), \notag \\
	\frac{\psi_{i+}}{\psi_{i-}} &\approx (2ik_i\epsilon)^{\zeta_i}, \quad \text{and} \notag \\
	Z &\approx \frac{\hat \psi_{2-}^\prime + R\,\hat \psi_{2+}^\prime}{\psi_{2-} + R\,\psi_{2+}} - h_{22} = \frac{\hat \psi_{2-}^\prime}{\psi_{2-}}\frac{1 + R\,\hat \psi_{2+}^\prime/\hat \psi_{2-}^\prime}{1 + R\,\psi_{2+}/\psi_{2-}} - h_{22} \notag \\
	  &\approx -\qty[(1 + \zeta_2)\frac{\pi\epsilon}{m_2}]\qty{1 + R\,\qty(\frac{\hat \psi_{2+}^\prime}{\hat \psi_{2-}^\prime} - \frac{\psi_{2+}}{\psi_{2-}})} - h_{22}, \notag \\
		&= -\qty[(1 + \zeta_2)\frac{\pi\epsilon}{m_2}]\qty{1 - \frac{2\zeta_2 R\,}{\zeta_2 + 1}(2ik_2\epsilon)^{\zeta_2}} - h_{22}.
\end{align}
Substituting in \eqref{eq:app:2PICs:genC1}, one finds
\begin{align}
	\label{eq:app:2PICs:C1NLO}
	C_{1-} &\approx -(2ik_1\epsilon)^{\zeta_1}\frac{\left\{\abs{h_{12}}^2 - \left[\frac{\pi\epsilon}{m_1} (1 - \zeta_1) + h_{11}\right]\qty[(1 + \zeta_2)\frac{\pi\epsilon}{m_2}\qty(1 - \frac{2\zeta_2 R\,}{\zeta_2 + 1}(2ik_2\epsilon)^{\zeta_2}) + h_{22}]\right\}}{\left\{\abs{h_{12}}^2 - \left[\frac{\pi\epsilon}{m_1} (1 + \zeta_1) + h_{11}\right]\qty[(1 + \zeta_2)\frac{\pi\epsilon}{m_2}\qty(1 - \frac{2\zeta_2 R\,}{\zeta_2 + 1}(2ik_2\epsilon)^{\zeta_2}) + h_{22}]\right\}}, \notag \\
		   &= -(2ik_1\epsilon)^{\zeta_1}\frac{\mathcal{N}_1 + \left[\frac{\pi\epsilon}{m_1} (1 - \zeta_1) + h_{11}\right]\frac{2\pi\epsilon\zeta_2 R\,}{m_2}(2ik_2\epsilon)^{\zeta_2}}{\mathcal{D} + \left[\frac{\pi\epsilon}{m_1} (1 + \zeta_1) + h_{11}\right]\frac{2\pi\epsilon\zeta_2 R\,}{m_2}(2ik_2\epsilon)^{\zeta_2}}, \notag \\
		   &\approx -(2ik_1\epsilon)^{\zeta_1}\frac{\mathcal{N}_1}{\mathcal{D}} \left\{ 1 + \left[ \frac{\left[\frac{\pi\epsilon}{m_1} (1 - \zeta_1) + h_{11}\right]}{\mathcal{N}_1} - \frac{\left[\frac{\pi\epsilon}{m_1} (1 + \zeta_1) + h_{11}\right]}{\mathcal{D}} \right]\frac{2\pi\epsilon\zeta_2 R\,}{m_2}(2ik_2\epsilon)^{\zeta_2} \right\},
\end{align}
where 
\begin{align}
	\label{eq:app:2PICs:defNandD}
	\mathcal{N}_1 &:= \abs{h_{12}}^2 - \left[\frac{\pi\epsilon}{m_1} (1 - \zeta_1) + h_{11}\right]\left[(1 + \zeta_2)\frac{\pi\epsilon}{m_2} + h_{22}\right], \quad \text{and} \notag \\
	\mathcal{D} &:= \abs{h_{12}}^2 - \left[\frac{\pi\epsilon}{m_1} (1 + \zeta_1) + h_{11}\right]\left[(1 + \zeta_2)\frac{\pi\epsilon}{m_2} + h_{22}\right].
\end{align}
Then \eqref{eq:app:2PICs:C1NLO} simplifies to:
\begin{align}
	\label{eq:app:2PICs:C1NLOSimple}
	C_{11} &= -(2ik_1\epsilon)^{\zeta_1}\frac{\widehat{\mathcal{N}}_1}{\widehat{\mathcal{D}}} \left\{ 1 - 4R\,\frac{\hat\lambda_{12}\hat\lambda_{21}}{\widehat{\mathcal{N}_1}\widehat{\mathcal{D}}}(2ik_2\epsilon)^{\zeta_2} \right\}.
\end{align}

\section{Solving for Point-Particle Couplings}
\label{appendix:solvePPs}

To find the running of the point-particle couplings, we need to isolate for them in the boundary conditions. To do so, we follow the same prescription as \ref{appendix:2PICs}. Write:
\begin{equation}
	\label{eq:app:PPs:genPsi}
	\psi_i = C_{i+} \psi_{i+} + C_{i-}\psi_{i-},
\end{equation}
where $i = 1,2$. We again use \eqref{eq:scattering:11:defA} to define $C_{11}:= C_{1-}/C_{1+}$ and $C_{12} := C_{2-}/C_{1+}$ in the $1\to X$ system, and analogously $C_{22}:= C_{2-}/C_{2+}$ and $C_{21} := C_{1-}/C_{2+}$ in the $2 \to X$ system. 

For convenience, define the following:
\begin{equation}
	\label{eq:app:PPs:sysA}
	\psi_{11} := \psi_{1+} + C_{11}\psi_{1-} \qquad \text{and} \qquad \psi_{12} := C_{12} \left[ R \psi_{2+} + \psi_{2-}\right]
\end{equation}
for the $1\to X$ system, and
\begin{equation}
	\label{eq:app:PPs:sysB}
	\psi_{22} := \psi_{2+} + C_{22}\psi_{2-} \qquad \text{and} \qquad \psi_{21} := C_{21} \left[ R \psi_{1+} + \psi_{1-}\right]
\end{equation}
for the $2\to X$ system. The small-$r$ boundary conditions \eqref{eq:BC:IC:rbcA} and \eqref{eq:BC:IC:rbcB} can be written
\begin{align}
	\hat\psi_{11}^\prime &= h_{11}\psi_{11} + h_{12}\psi_{12}, \quad \text{and} \label{eq:app:PPs:genBCAa} \\
	\hat\psi_{12}^\prime &= h_{22}\psi_{12} + h_{21}\psi_{11}, \label{eq:app:PPs:genBCAb}
\end{align}
and 
\begin{align}
	\hat\psi_{21}^\prime &= h_{11}\psi_{21} + h_{12}\psi_{22}, \quad \text{and} \label{eq:app:PPs:genBCBa}\\
	\hat\psi_{22}^\prime &= h_{22}\psi_{22} + h_{21}\psi_{21}, \label{eq:app:PPs:genBCBb}
\end{align}
As in appendix \ref{appendix:2PICs}, we define $\hat \psi^\prime := 4\pi\epsilon^2\partial_\epsilon\psi$. Using \eqref{eq:app:PPs:genBCAa} and \eqref{eq:app:PPs:genBCBa}, we can isolate for $h_{11}$ and $h_{12}$:
\begin{equation}
	\label{eq:app:PPs:isolM1X}
	h_{11} = \frac{\hat\psi_{11}^\prime\psi_{22} - \hat\psi_{21}^\prime\psi_{12}}{\psi_{11}\psi_{22} - \psi_{21}\psi_{12}} \qquad \text{and} \qquad h_{12} = \frac{\hat\psi_{11}^\prime\psi_{21} - \hat\psi_{21}^\prime\psi_{11}}{\psi_{12}\psi_{21} - \psi_{22}\psi_{11}}.
\end{equation}
Similarly, using \eqref{eq:app:PPs:genBCAb} and \eqref{eq:app:PPs:genBCBb}, we can isolate for $h_{22}$ and $h_{21}$:
\begin{equation}
	\label{eq:app:PPs:isolM2X}
	h_{22} = \frac{\hat\psi_{12}^\prime\psi_{21} - \hat\psi_{22}^\prime\psi_{11}}{\psi_{12}\psi_{21} - \psi_{22}\psi_{11}} \qquad \text{and} \qquad h_{21} = \frac{\hat\psi_{12}^\prime\psi_{22} - \hat\psi_{22}^\prime\psi_{12}}{\psi_{11}\psi_{22} - \psi_{21}\psi_{12}}.
\end{equation}

To make use of these formulae, we approximate $\psi_{ij}$ using the small-$r$ forms as used in appendix \ref{appendix:2PICs}:
\begin{align}
	\label{eq:app:PPs:smallrPsi1LO}
	\psi_{11(22)}(\epsilon) &\approx x_{1(2)}^{-1/2}\qty[x_{1(2)}^{\zeta_{1(2)}/2} + C_{11(22)}x_{1(2)}^{-\zeta_{1(2)}/2}], \quad \text{and} \notag \\
	4\pi\epsilon^2\partial_r\vert_{\epsilon} \psi_{11(22)}(\epsilon) &\approx 2\pi\epsilon\, x_{1(2)}^{-1/2}\qty[(\zeta_{1(2)} - 1)x_{1(2)}^{\zeta_{1(2)}/2} - (\zeta_{1(2)} + 1)C_{11(22)}x_{1(2)}^{-\zeta_{1(2)}/2}],
\end{align}
and
\begin{align}
	\label{eq:app:PPs:smallrPsi2LO}
	\psi_{21(12)}(\epsilon) &\approx C_{21(12)}x_{1(2)}^{-1/2 - \zeta_{1(2)}/2}, \quad \text{and} \notag \\
	4\pi\epsilon^2\partial_r\vert_\epsilon \psi_{21(12)}(\epsilon) &\approx - (\zeta_{1(2)} + 1)2\pi\epsilon\, C_{21(12)} x_{1(2)}^{-1/2 - \zeta_{1(2)}/2}.
\end{align}
where again, $x_{i} := (2ik_i\epsilon)$, with $i = 1,2$. Substituting \eqref{eq:app:PPs:smallrPsi1LO} and \eqref{eq:app:PPs:smallrPsi2LO} into \eqref{eq:app:PPs:isolM1X} and \eqref{eq:app:PPs:isolM2X}, we have the following. For $h_{11}$ we find
\begin{align}
	\label{eq:app:PPs:solveM11}
	\hat \lambda_{11} &=  \frac{(x_1^{\zeta_1/2} - C_{11}x_1^{-\zeta_2/2})(x_2^{\zeta_1/2} + C_{22}x_2^{-\zeta_2/2}) + C_{21}C_{12}x_1^{-\zeta_1/2}x_2^{-\zeta_2/2}}{(x_1^{\zeta_1/2} + C_{11}x_1^{-\zeta_2/2})(x_2^{\zeta_1/2} + C_{22}x_2^{-\zeta_2/2}) - C_{21}C_{12}x_1^{-\zeta_1/2}x_2^{-\zeta_2/2}}, \notag \\
					&=  \frac{(1 - C_{11}x_1^{-\zeta_1})(1 + C_{22}x_2^{-\zeta_2}) + C_{21}C_{12}x_1^{-\zeta_1}x_{2}^{-\zeta_2} }{(1 + C_{11}x_1^{-\zeta_1})(1 + C_{22}x_2^{-\zeta_2}) - C_{21}C_{12}x_1^{-\zeta_1}x_{2}^{-\zeta_2} } ,
\end{align}
which defines $\hat{\lambda}_{11} := \frac{1}{\zeta_1}\left(\frac{m_1h_{11}}{\pi\epsilon} + 1\right)$. Notice the limit $C_{21} = C_{12} \to 0$ reduces $\hat \lambda_{11}$ to the single-species running \eqref{eq:review:run}, as it should (the limit in which there is no mixing between particle species). For $h_{12}$, we have
\begin{align}
	\label{eq:app:PPs:solveM12}
	\hat \lambda_{12} &=  \sqrt{\frac{m_2\zeta_1}{m_1\zeta_2}}\frac{C_{21}\left(\frac{x_{1}}{x_{2}}\right)^{-1/2}}{(x_1^{\zeta_1/2} + C_{11}x_1^{-\zeta_2/2})(x_2^{\zeta_1/2} + C_{22}x_2^{-\zeta_2/2}) - C_{21}C_{12}x_1^{-\zeta_1/2}x_2^{-\zeta_2/2}} , \notag \\
					&=  \sqrt{\frac{m_2\zeta_1}{m_1\zeta_2}} \frac{C_{21}\left(\frac{x_{1}}{x_{2}}\right)^{-1/2}x_1^{-\zeta_1/2}x_2^{-\zeta_2/2}}{(1 + C_{11}x_1^{-\zeta_1})(1 + C_{22}x_2^{-\zeta_2}) - C_{21}C_{12}x_1^{-\zeta_1}x_{2}^{-\zeta_2}} ,
\end{align}
with now $\hat{\lambda}_{12} := \frac{h_{12}\sqrt{m_1m_2} }{2\pi\epsilon \sqrt{\zeta_1\zeta_2}}$. Notice again the clean limit $h_{12} \to 0$ when $C_{21} \to 0$. The rest follow easily:
\begin{align}
	\label{eq:app:PPs:solveM21}
	\hat\lambda_{21} &= \sqrt{\frac{m_1\zeta_2}{m_2\zeta_1}} \frac{C_{12}\left(\frac{x_{2}}{x_{1}}\right)^{-1/2}x_1^{-\zeta_1/2}x_2^{-\zeta_2/2}}{(1 + C_{11}x_1^{-\zeta_1})(1 + C_{22}x_2^{-\zeta_2}) - C_{21}C_{12}x_1^{-\zeta_1}x_{2}^{-\zeta_2}} ,
\end{align}
similarly with $\hat{\lambda}_{21} := \frac{h_{21}\sqrt{m_1 m_2}}{2\pi\epsilon \sqrt{\zeta_1\zeta_2}}$. Lastly, 
\begin{align}
	\label{eq:app:PPs:solveM22}
	\hat \lambda_{22} &=  \frac{(1 + C_{11}x_1^{-\zeta_1})(1 - C_{22}x_2^{-\zeta_2}) + C_{21}C_{12}x_1^{-\zeta_1}x_{2}^{-\zeta_2} }{(1 + C_{11}x_1^{-\zeta_1})(1 + C_{22}x_2^{-\zeta_2}) - C_{21}C_{12}x_1^{-\zeta_1}x_{2}^{-\zeta_2} } ,
\end{align}
with $\hat{\lambda}_{22} := \frac{1}{\zeta_2}\left(\frac{m_2h_{22}}{2\pi\epsilon} + 1\right)$.


\bibliographystyle{JHEP}
\bibliography{refs}

\end{document}